\documentclass[iop,twocolappendix]{emulateapj}

\usepackage{hyperref}
\usepackage{multirow}
\usepackage{subfigure}
\usepackage{natbib}
\usepackage{graphicx}
\usepackage{amsmath}

\def \mic{~$\mu$m}

\slugcomment{Draft: \today}

\shorttitle{Palomar/TripleSpec observations of {\it Spitzer}/MIPSGAL~24\mic\ circumstellar shells}
\shortauthors{Flagey et al.}

\begin{document}

\title{Palomar/TripleSpec observations of {\it Spitzer}/MIPSGAL~24\mic\ circumstellar shells: unveiling the nature of their central sources}

\author{N. Flagey\altaffilmark{1,4}}
\author{A. Noriega-Crespo\altaffilmark{2,5}}
\author{A. Petric\altaffilmark{3,6}}
\author{T.~R. Geballe\altaffilmark{6}}

\email{nflagey@jpl.nasa.gov}

\altaffiltext{1}{Jet Propulsion Laboratory, California Institute of Technology, 4800 Oak Grove Drive, Pasadena, CA 91109, USA}
\altaffiltext{2}{Infrared Processing and Analysis Center, California Institute of Technology, 1200 East California Blvd, Pasadena, CA 91125, USA}
\altaffiltext{3}{California Institute of Technology, 1200 East California Blvd, Pasadena, CA 91125, USA}
\altaffiltext{4}{Institute for Astronomy, 640 North A'ohoku Place, Hilo, HI 96720-2700, USA}
\altaffiltext{5}{Space Telescope Science Institute, 3700 San Martin Dr, Baltimore, MD 21218, USA}
\altaffiltext{6}{Gemini North Observatory, 670 North A'ohoku Place, Hilo, HI 96720, USA}

\begin{abstract}
We present near-IR spectroscopic observations of the central sources in 17 circumstellar shells from a sample of more than 400 ``bubbles'' discovered in the {\it Spitzer}/MIPSGAL 24~$\mu$m survey of the Galactic plane and in the Cygnus-X region. To identify the natures of these shells, we have obtained {\it J}, {\it H}, and {\it K} band spectra with a resolution $\sim$2600 of the stars at their centers. We observed 14 MIPSGAL bubbles (MBs), WR149, and 2 objects in the Cygnus-X region (WR138a and BD+43~3710), our sample being about 2.5 magnitudes fainter in K band than previous studies of the central sources of MBs. We use spectroscopic diagnostics and spectral libraries of late and early type stars to constrain the natures of our targets. We find five late type giants. The equivalent widths of their CO 2.29\mic\ features allow us to determine the spectral types of the stars and hence derive extinction along the line of sight, distance, and physical size of the shells. We also find twelve early type stars, in nine MBs and the 3 comparison objects. We find that the subtype inferred from the near-IR for WR138a (WN9h) and WR149 (WN5h) agrees with that derived from optical observations. A careful analysis of the literature and the environment of BD+43~3710 allows us to rule out the carbon star interpretation previously suggested. Our near-IR spectrum suggests that it is a B5 supergiant. At the centers of the nine MBs, we find a WC5-6 star possibly of low-mass, a candidate O5-6~V star, a B0 supergiant, a B/A type giant, and five LBV candidates. We also report the detections of emission lines arising from at least two shells with typical extents ($\sim10$\arcsec) in agreement with those in the mid-IR. We summarize the findings on the natures of the MBs since their discovery, with 30\% of them now known. Most MBs with central sources detected in the near- to mid-IR have been identified and are red and blue giants, supergiants, or stars evolving toward these phases including, in particular, a handful of newly discovered Wolf-Rayet stars and a significant number of LBV candidates.
\end{abstract}

\section{Introduction}

\begin{figure*}[!t]
  \centering
  \subfigure[MB3157]
  {\label{fig:mb3157}
    \includegraphics[width=.225\linewidth]{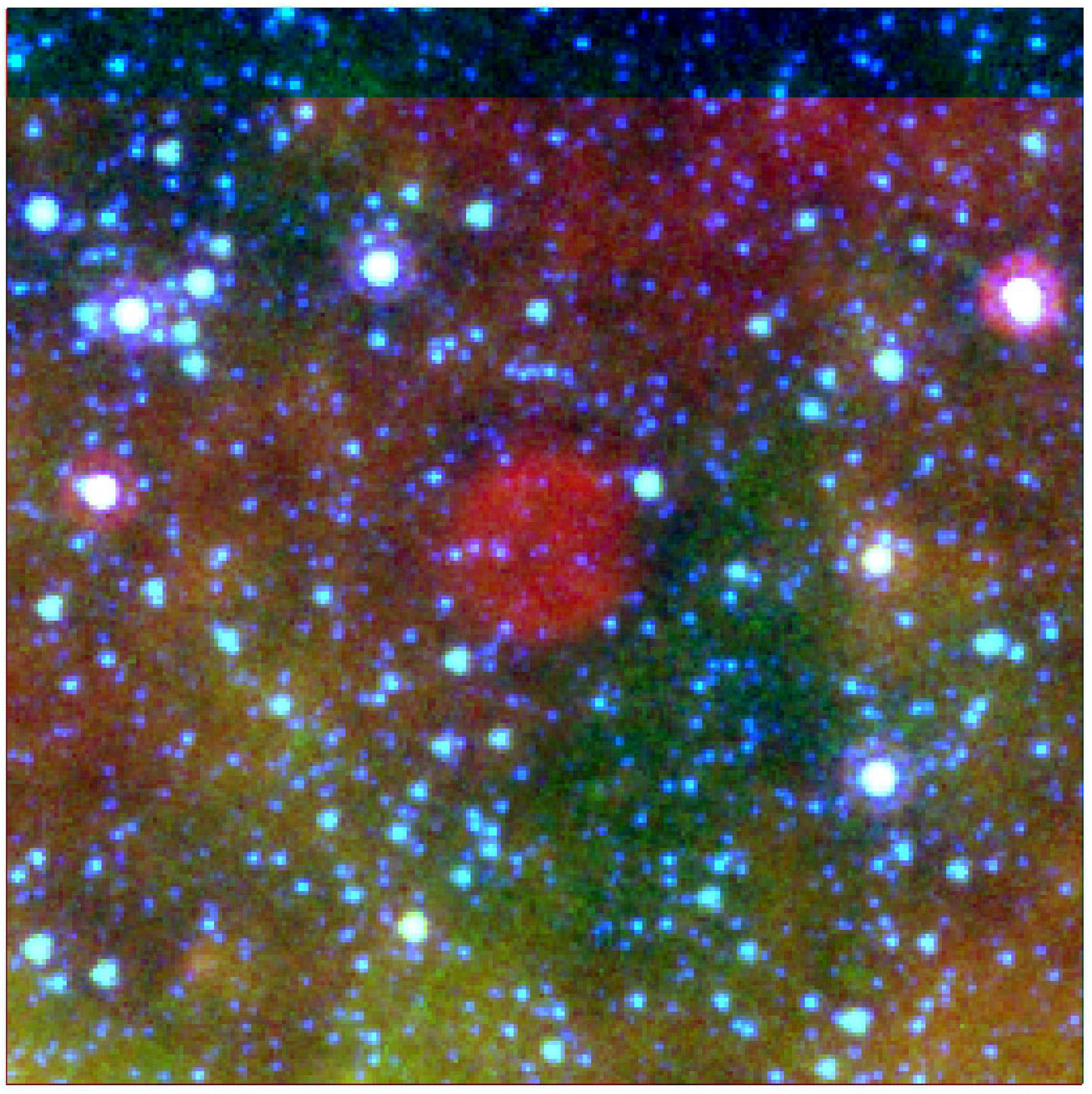}}
  \subfigure[MB3214]
  {\label{fig:mb3214}
    \includegraphics[width=.225\linewidth]{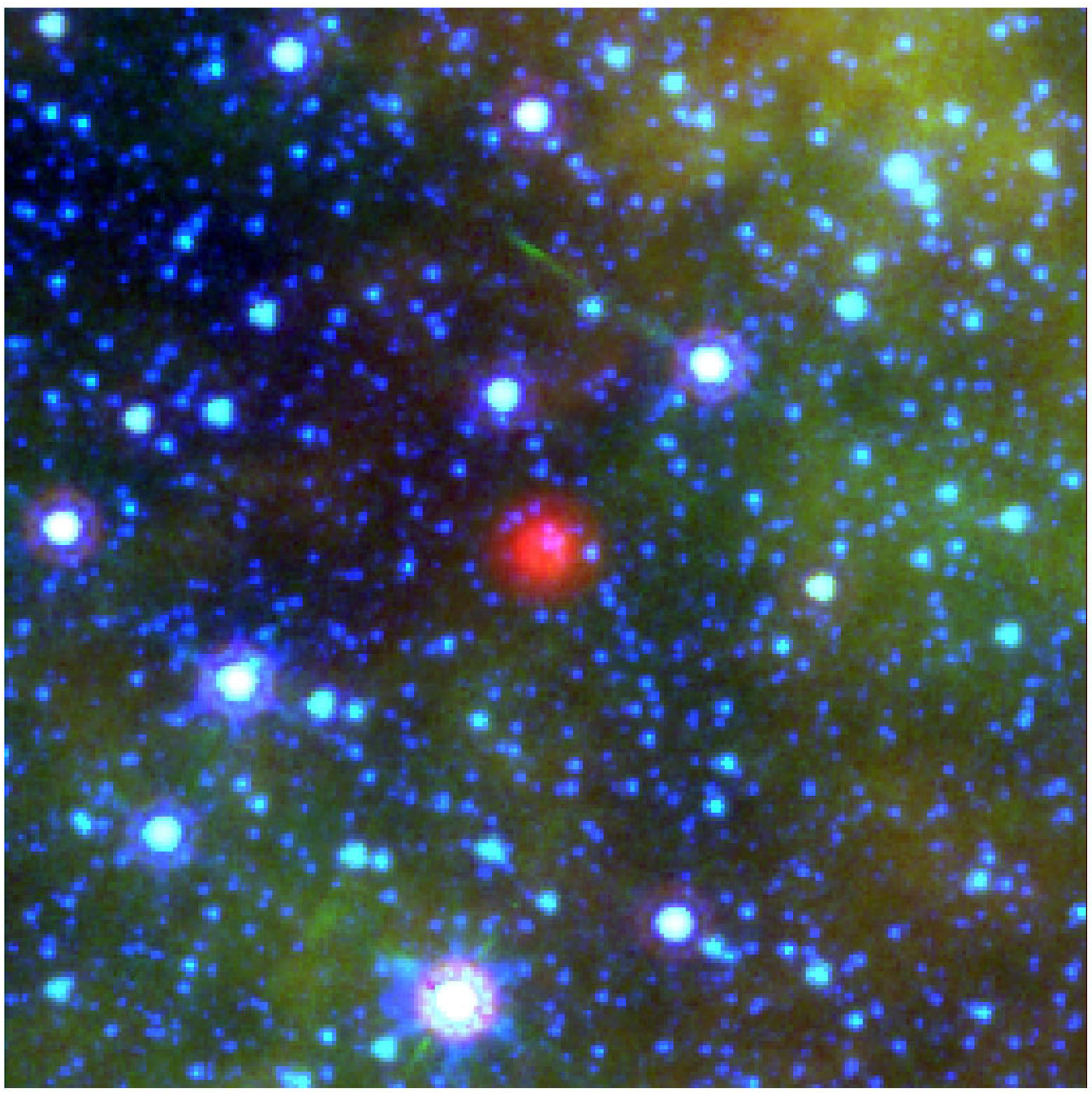}}
  \subfigure[MB3222]
  {\label{fig:mb3222}
    \includegraphics[width=.225\linewidth]{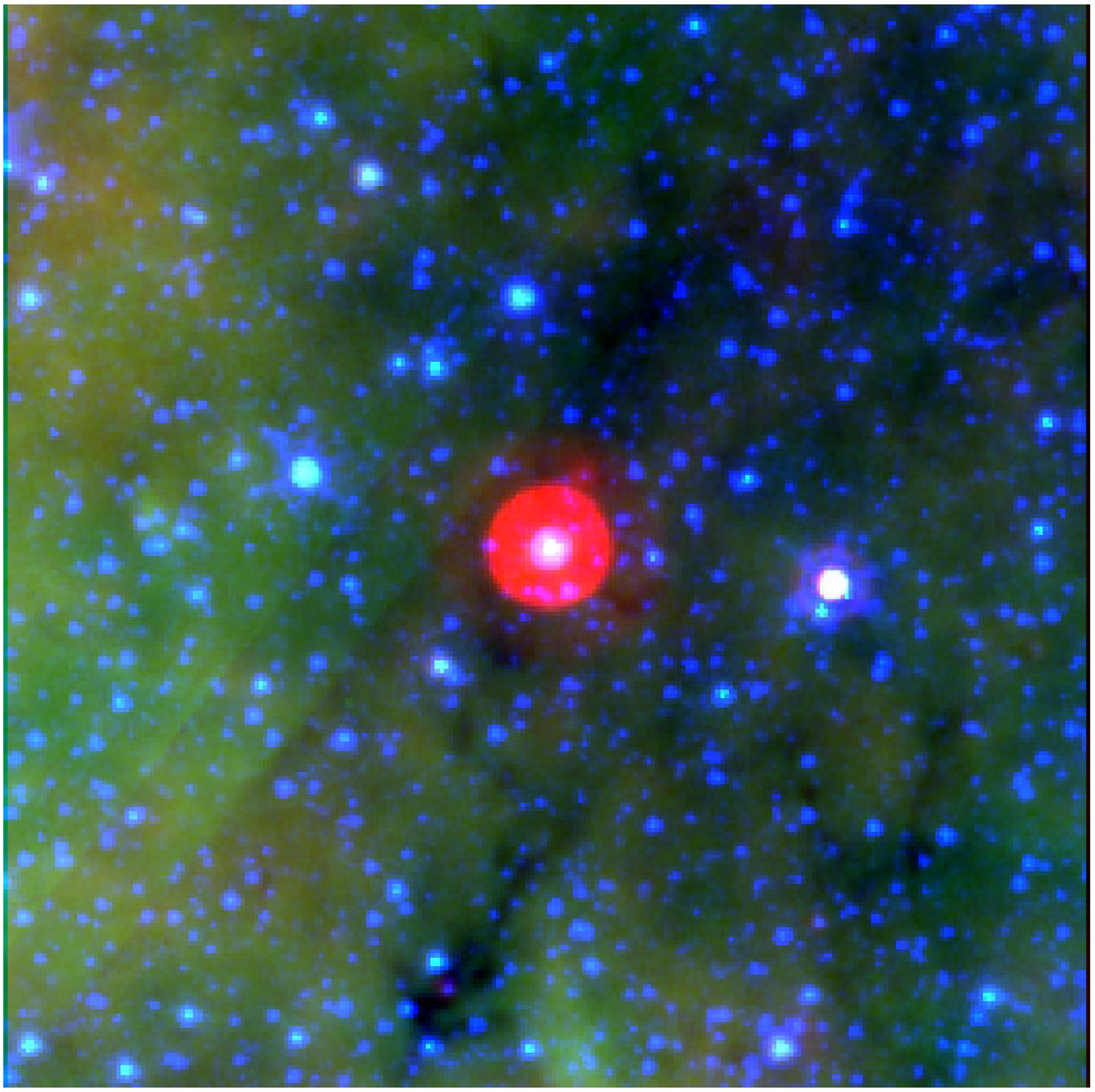}}
  \subfigure[MB3259]
  {\label{fig:mb3259}
    \includegraphics[width=.225\linewidth]{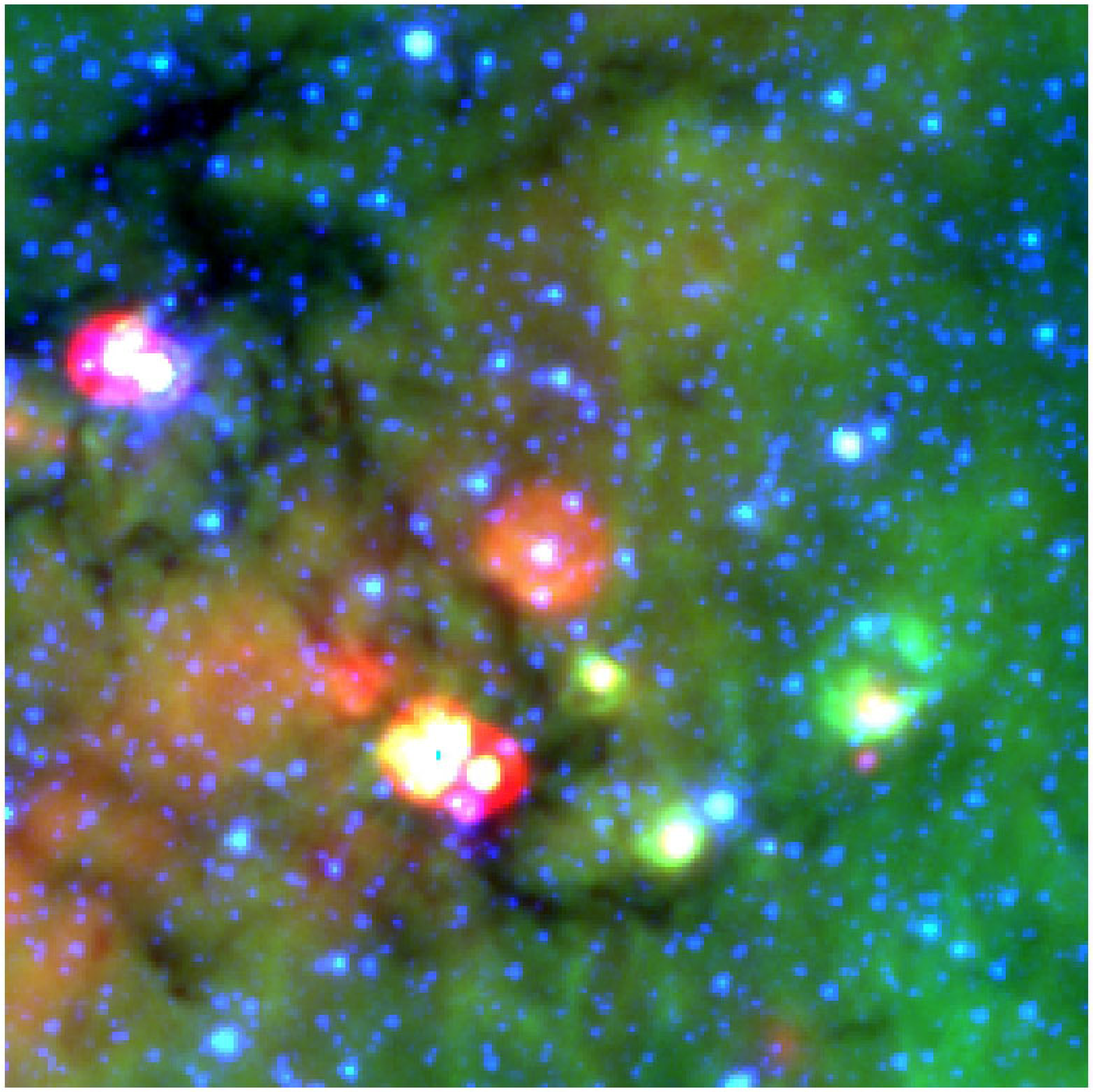}}
  \subfigure[MB3367]
  {\label{fig:mb3367}
    \includegraphics[width=.225\linewidth]{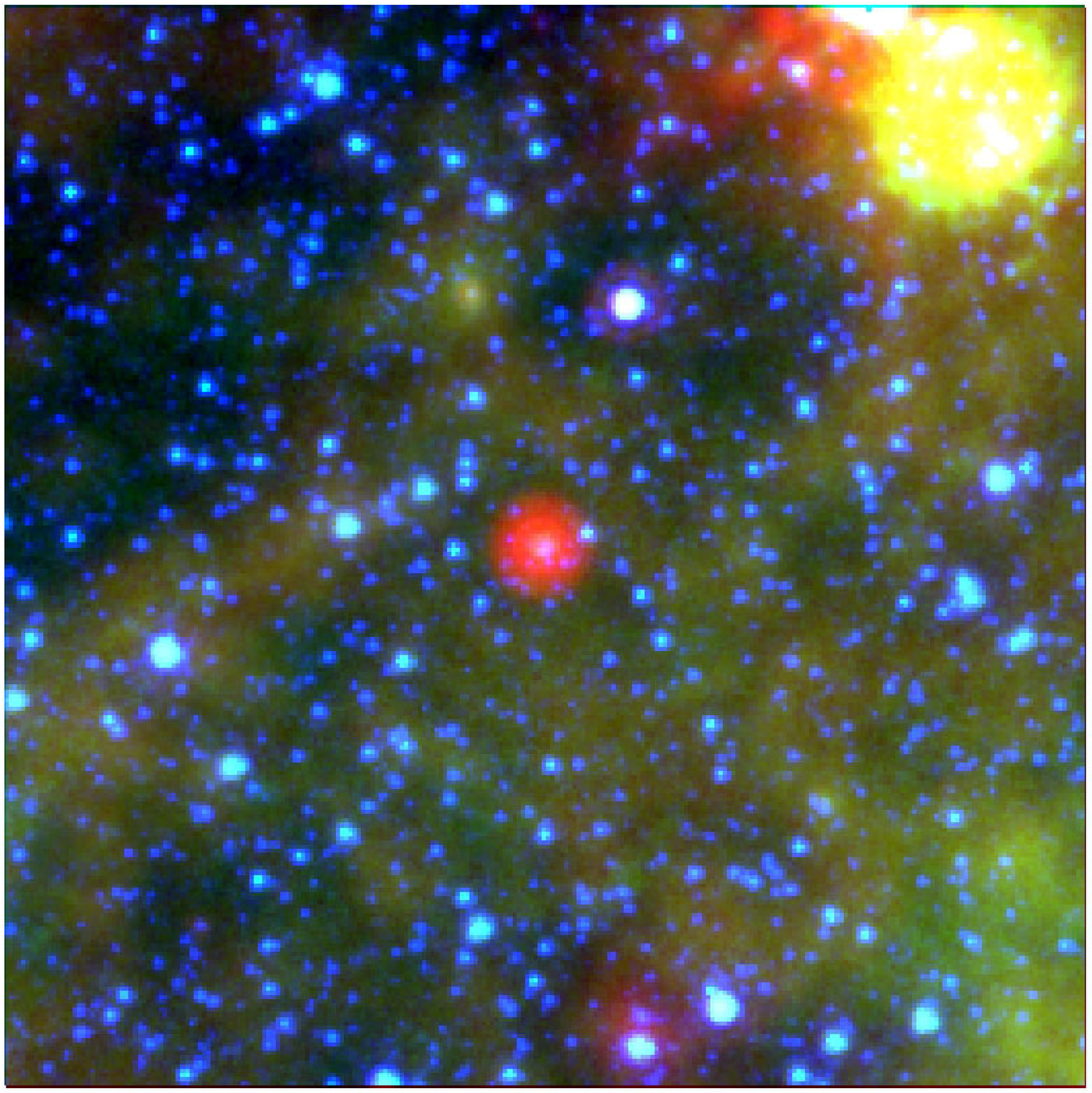}}
  \subfigure[MB3438]
  {\label{fig:mb3438}
    \includegraphics[width=.225\linewidth]{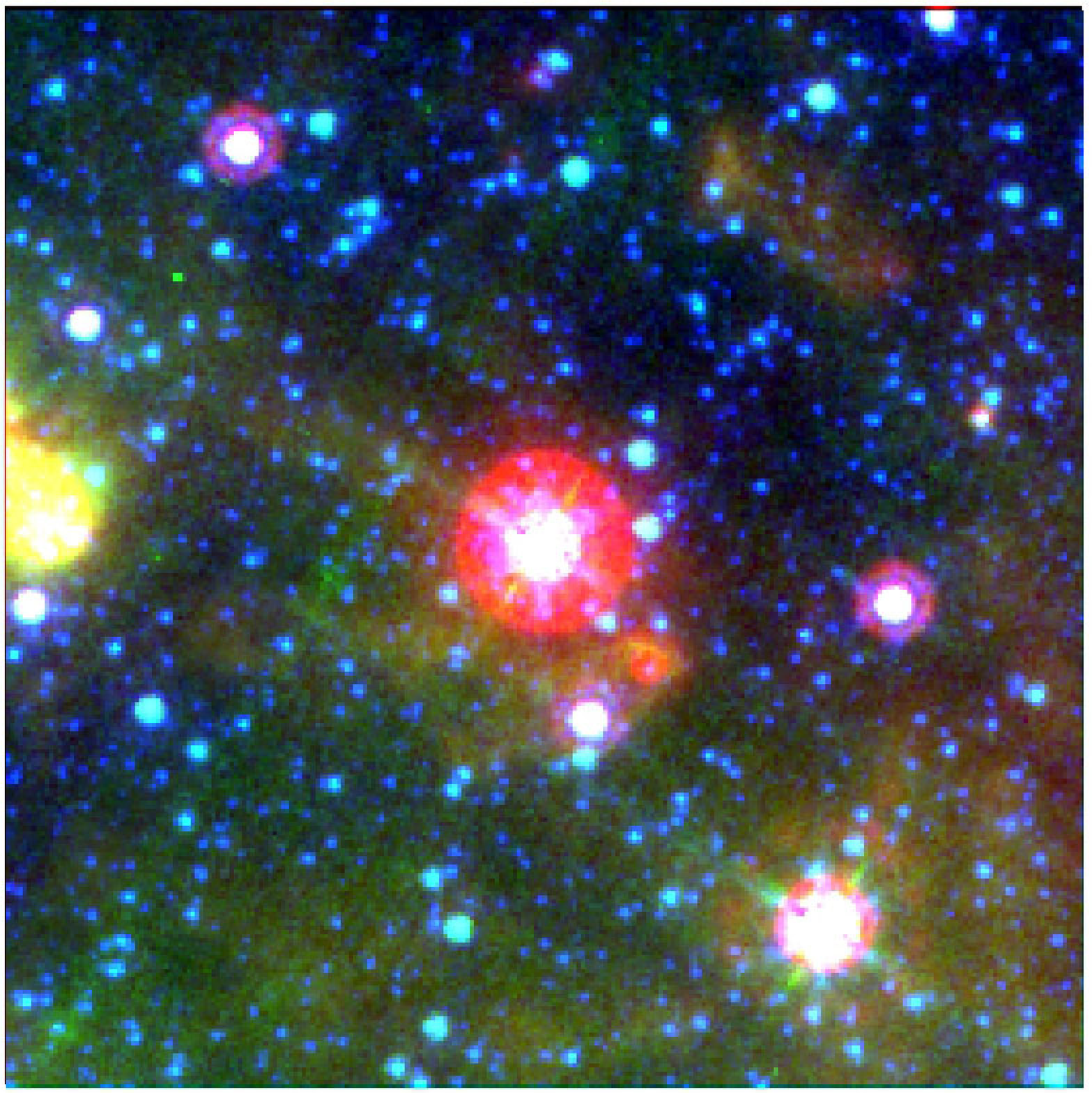}}
  \subfigure[MB3448]
  {\label{fig:mb3448}
    \includegraphics[width=.225\linewidth]{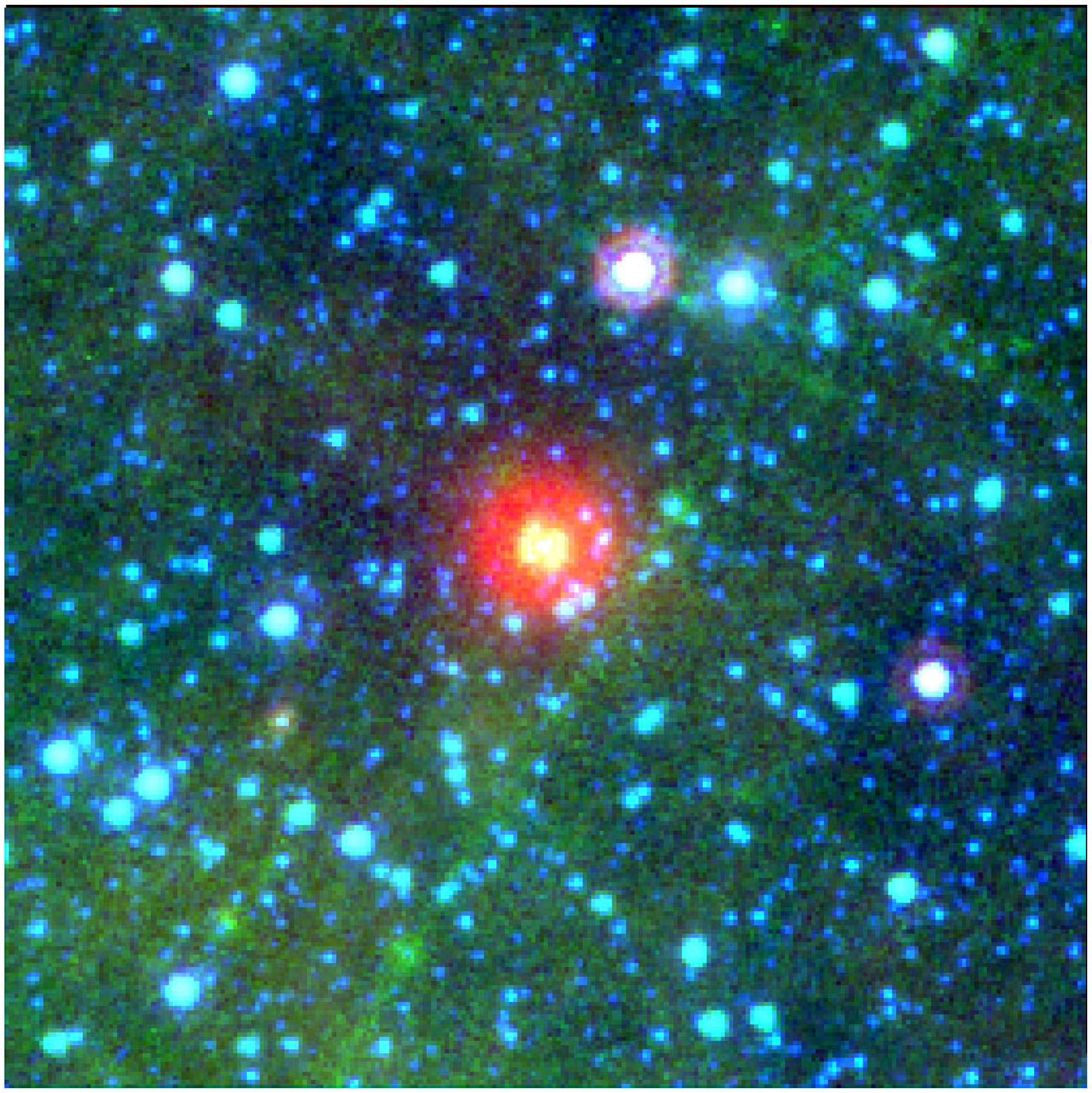}}
  \subfigure[MB3458]
  {\label{fig:mb3458}
    \includegraphics[width=.225\linewidth]{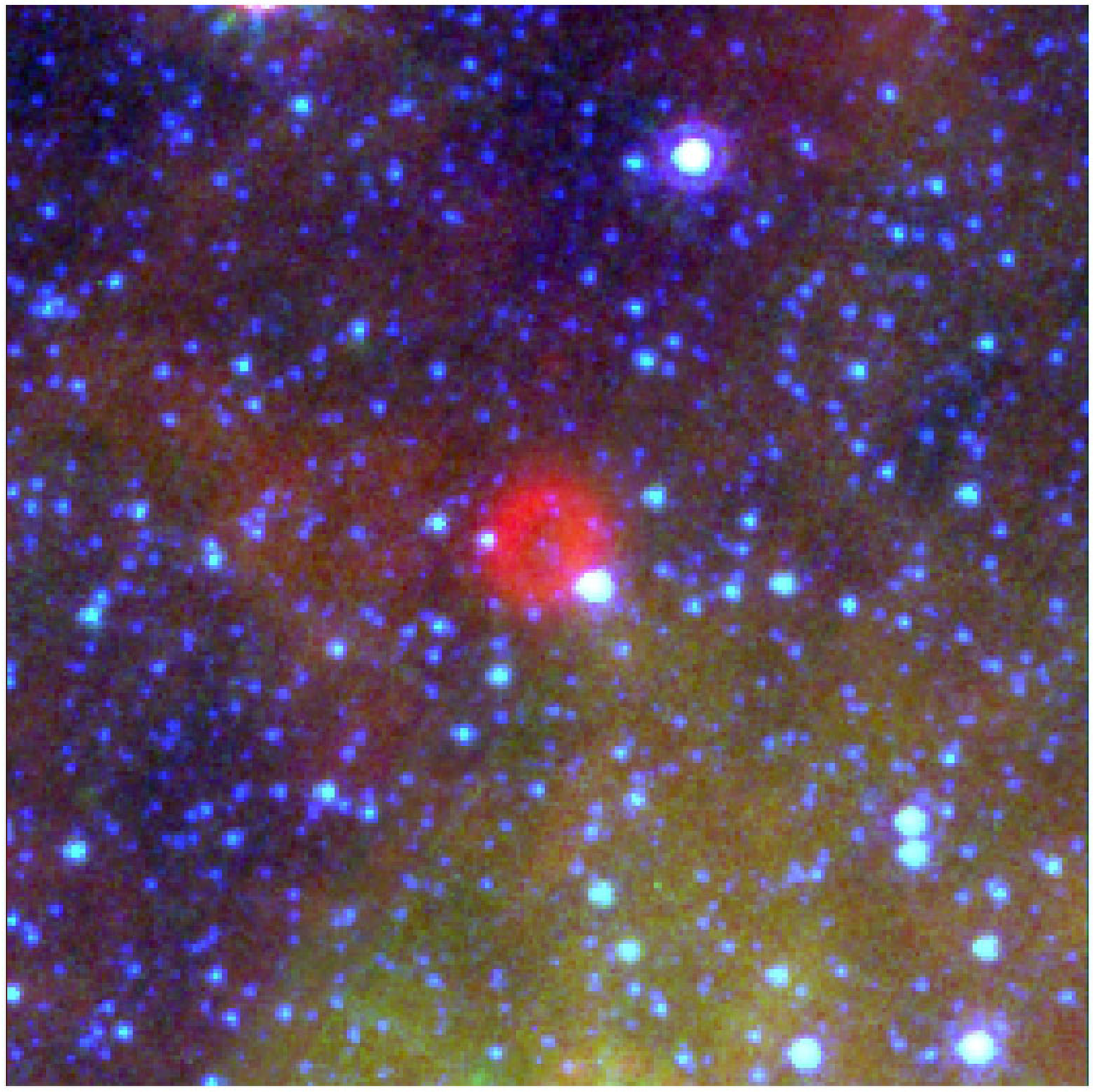}}
  \subfigure[MB3611]
  {\label{fig:mb3611}
    \includegraphics[width=.225\linewidth]{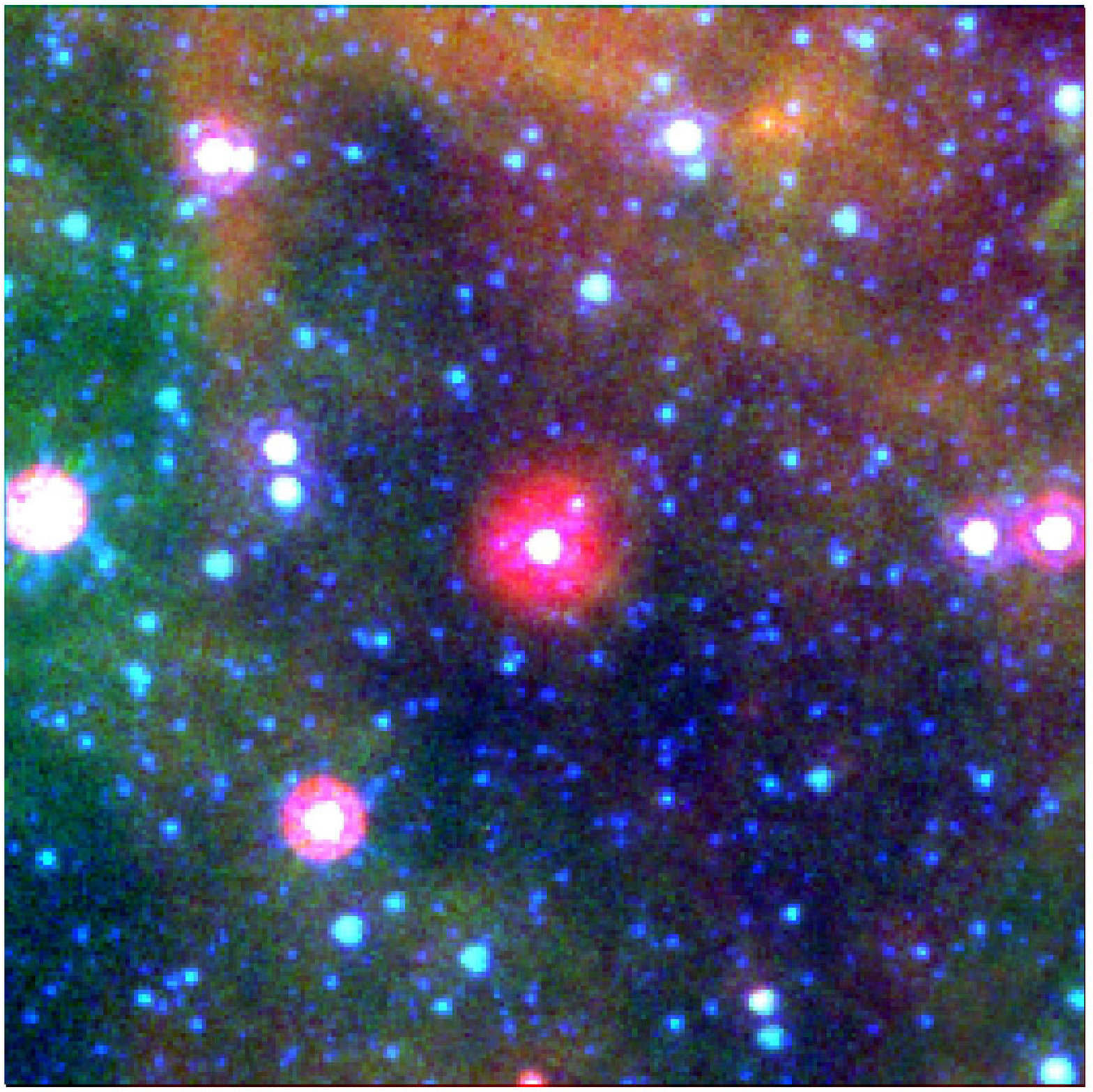}}
  \subfigure[MB3643]
  {\label{fig:mb3643}
    \includegraphics[width=.225\linewidth]{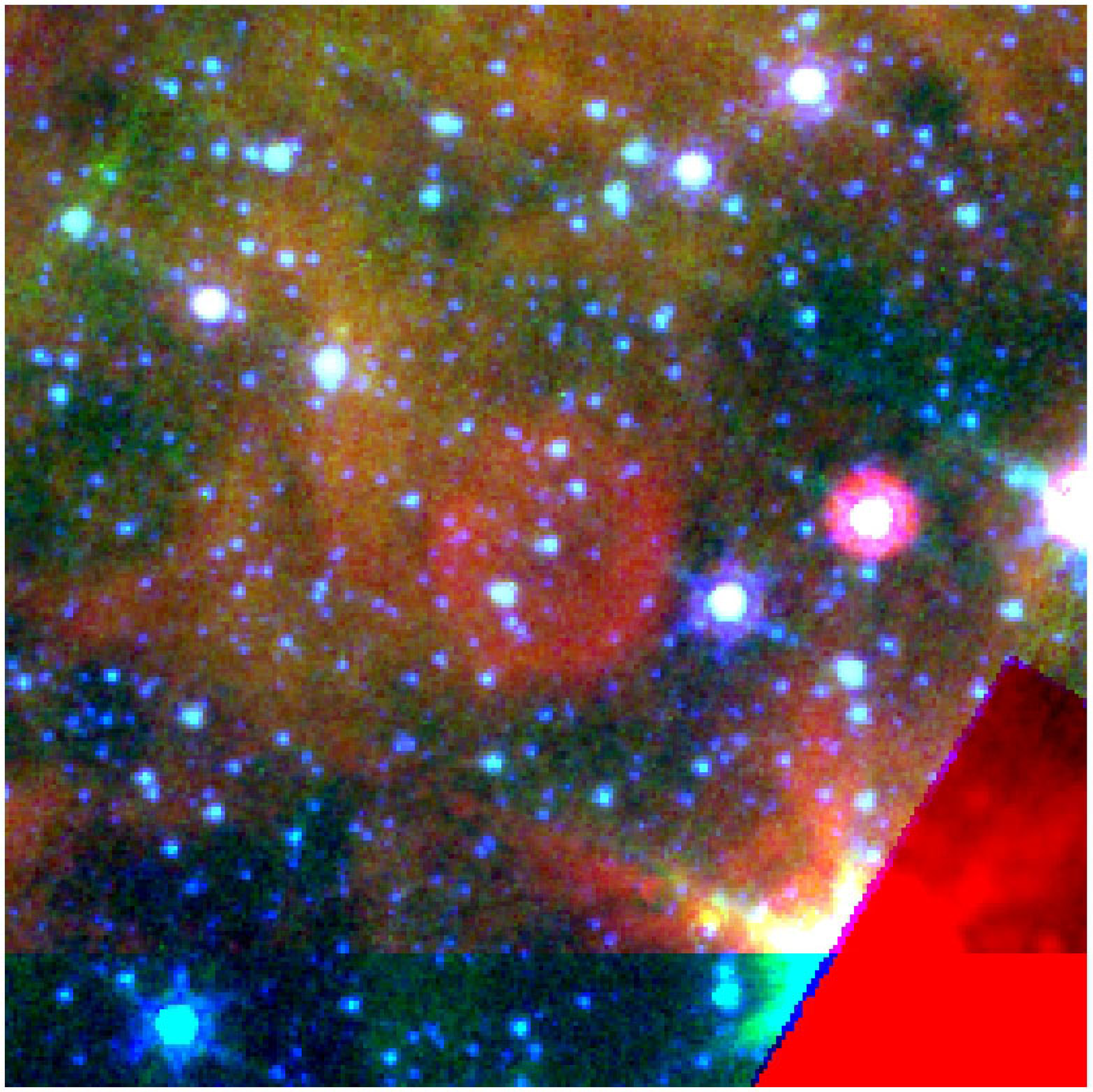}}
  \subfigure[MB3662]
  {\label{fig:mb3662}
    \includegraphics[width=.225\linewidth]{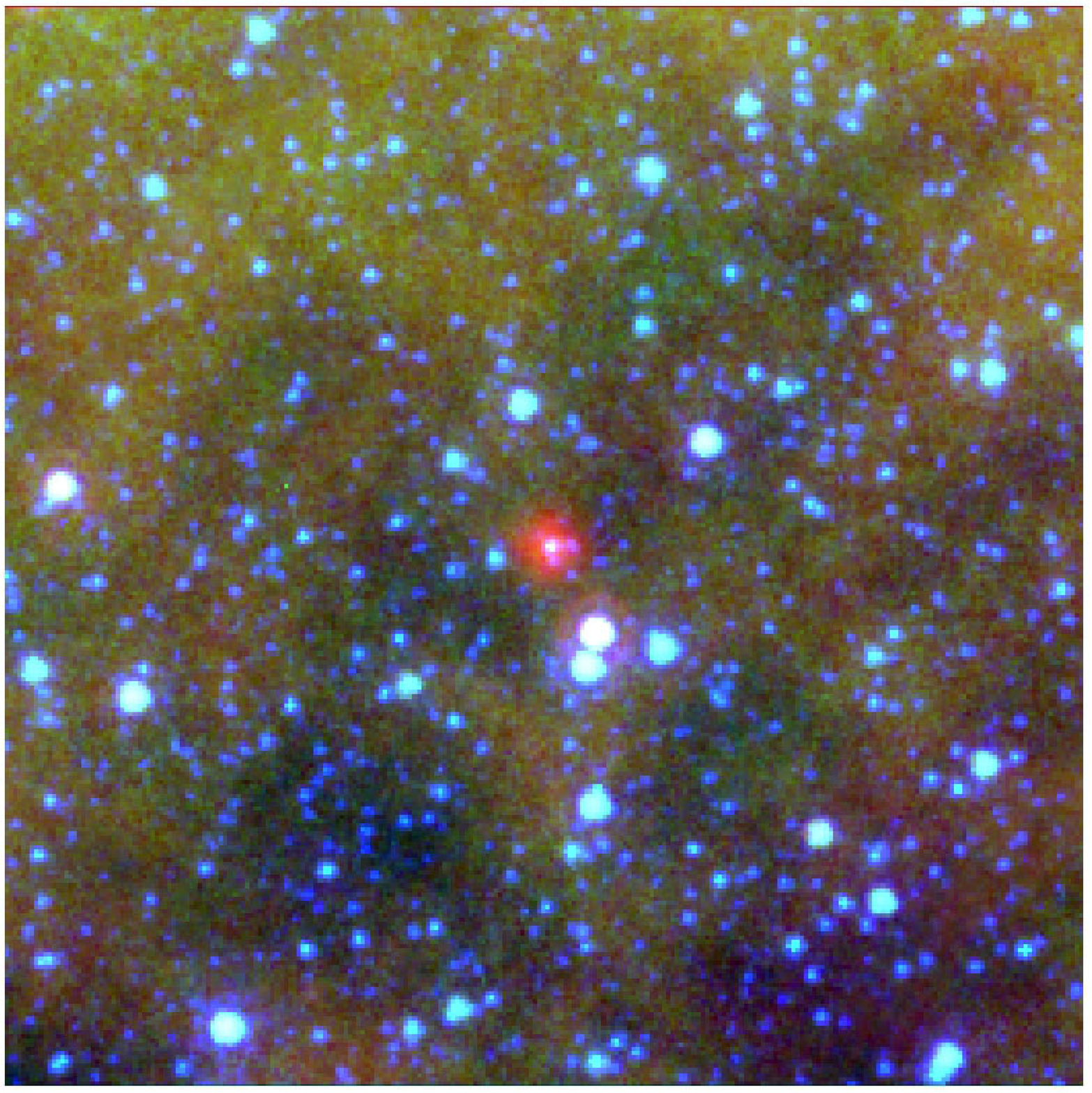}}
  \subfigure[MB3671]
  {\label{fig:mb3671}
    \includegraphics[width=.225\linewidth]{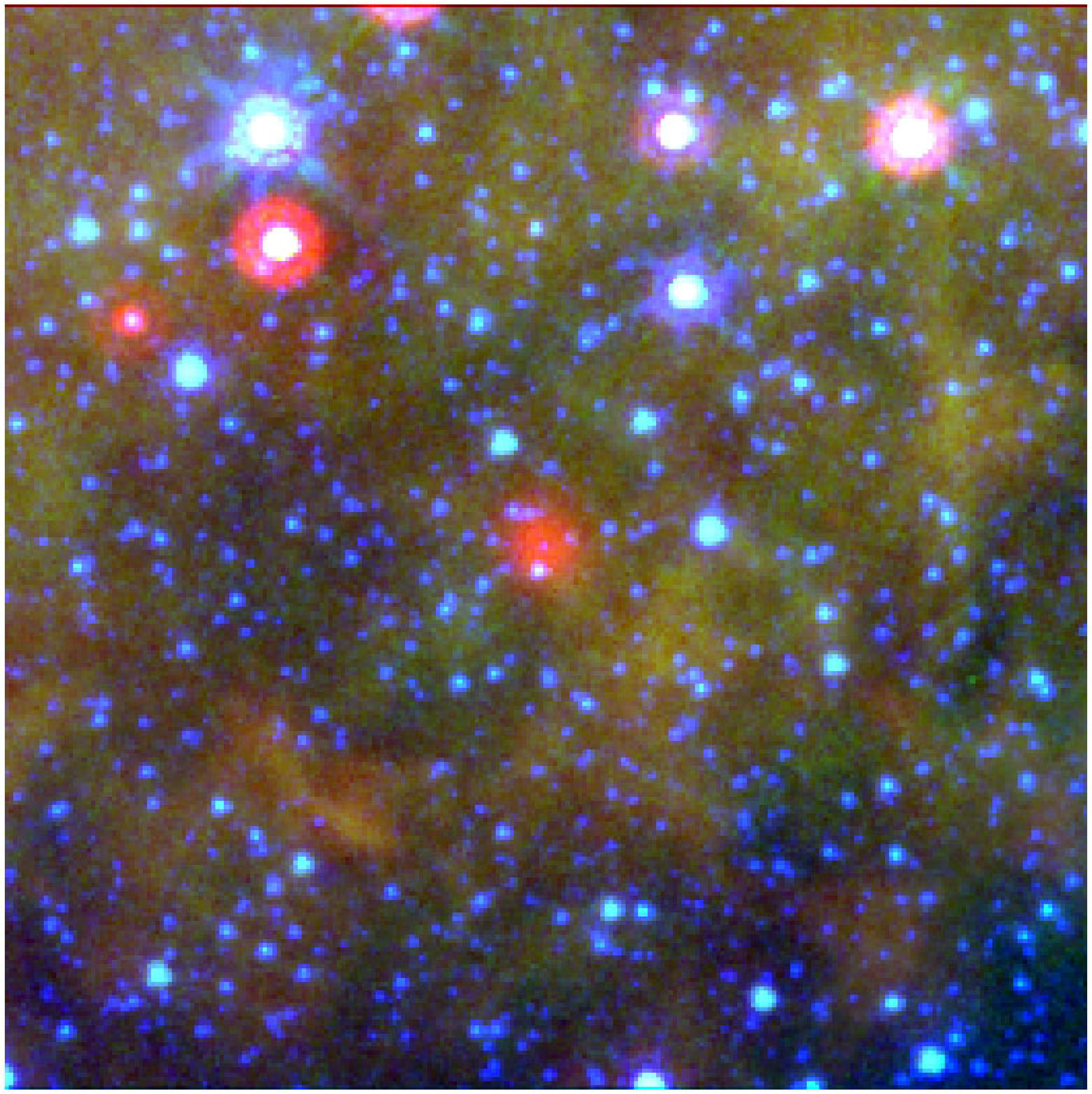}}
  \subfigure[MB3701]
  {\label{fig:mb3701}
    \includegraphics[width=.225\linewidth]{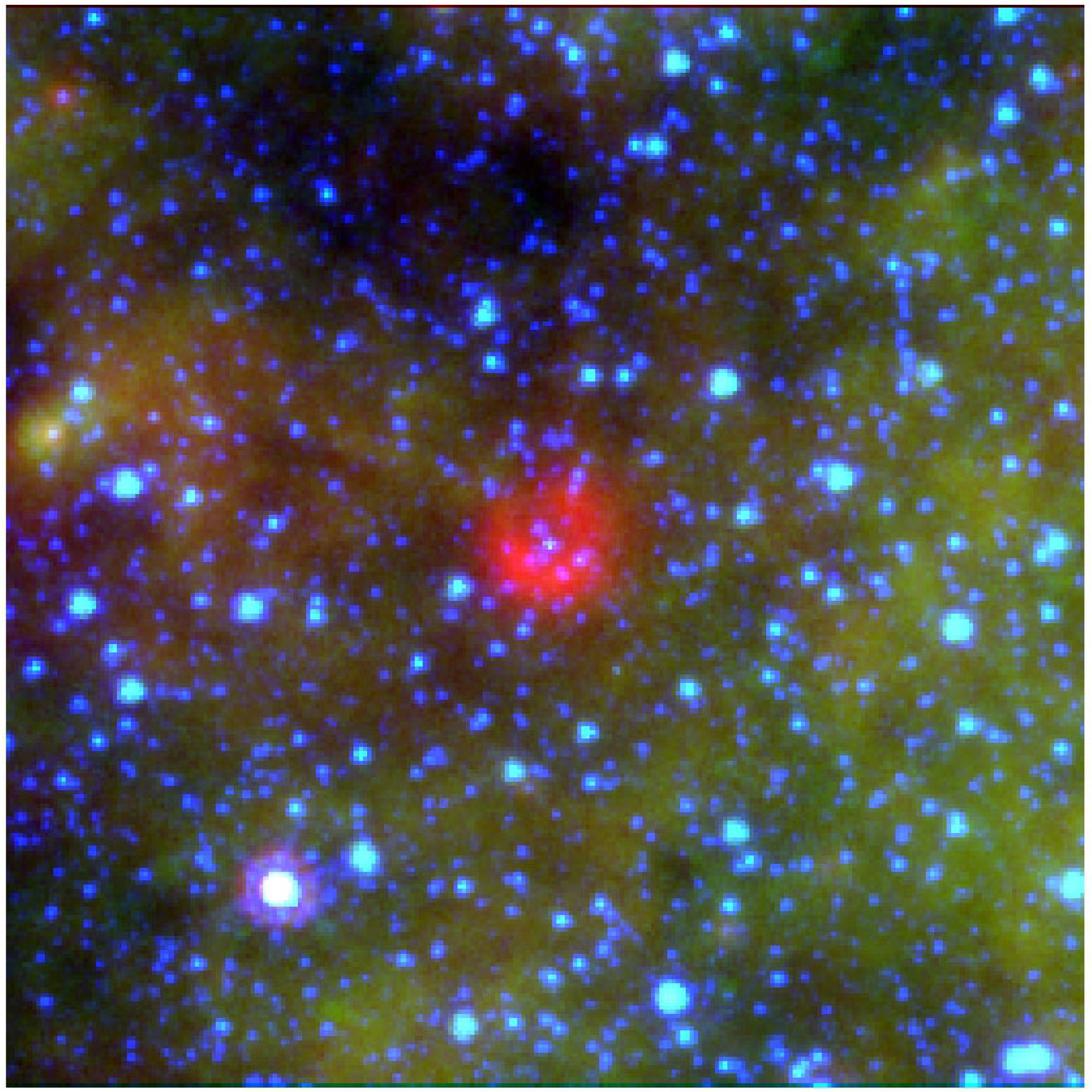}}
  \subfigure[MB3834]
  {\label{fig:mb3834}
    \includegraphics[width=.225\linewidth]{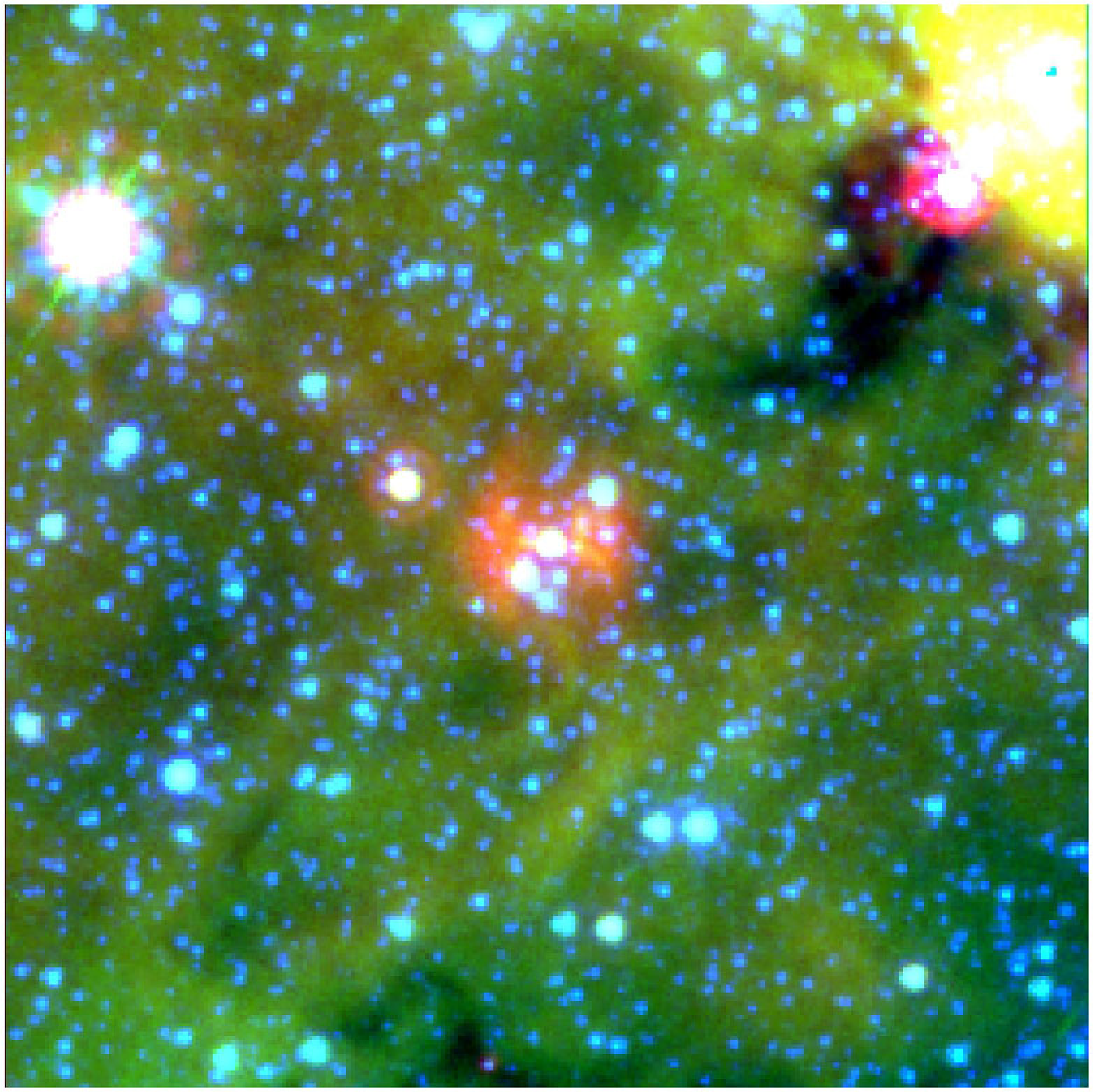}}
  \caption{Three color images (red is MIPS24, green is IRAC8, blue is IRAC4~$\mu$m) of all the targets from \citet{Mizuno2010} in our sample. Each image is 5\arcmin by 5\arcmin. The color scale of each band differs in each image so to enhance the local structure.}
  \label{fig:rgb}
\end{figure*}

Over 400 small ($\lesssim$1\arcmin) rings, disks and shells have been discovered from visual inspection of the {\it Spitzer}/MIPSGAL~24~$\mu$m mosaic images \citep{Carey2009, Mizuno2010}. These MIPSGAL bubbles (MBs) are pervasive through the entire Galactic plane in the mid-infrared. They span a large range of morphologies, sizes and fluxes. All the MBs have also been observed with the IRAC instrument on board {\it Spitzer}, as part of the GLIMPSE survey \citep{Benjamin2003}. The GLIMPSE images indicate that only 10\% of the MBs are detected between 3 and 8~$\mu$m. The analysis of the broadband images has also unveiled that 54 MBs (about 13\%) have central sources at 24~$\mu$m. This number rises to at least 100 in the IRAC or 2MASS images \citep[{\it J}, {\it H}, and {\it K} bands,][]{Skrutskie2006}. Several objects similar to the MBs have also been found in the Cygnus-X region, as reported by \citet{Gvaramadze2010}, \citet{Kraemer2010}, and \citet{Gvaramadze2012}. \citet{Gvaramadze2010} and \citet{Wachter2010} reported the discovery of several tens of similar objects in addition to those of \citet{Mizuno2010} in the Galactic plane and the Cygnus-X region.

When \citet{Mizuno2010} published their catalog, only 15\% of the MBs were identified or associated with stars of known spectral types. A large majority of the known MBs were found in the MASH Catalogue of planetary nebulae \citep[PNe, ][]{Parker2006} and in the Catalogue of Galactic PNe \citep{Kohoutek2001}. A few were associated with supernova remnants, Wolf-Rayet stars (WR), luminous blue variables (LBV) and other emission line stars. Most MBs were thus suspected to be associated with stars in late stages of evolution, with at least a fraction of them being massive. The extended mid-IR emission is expected to arise from warm, small dust grains and/or from hot ionized gas in the stellar winds and ejecta \citep{Barniske2008}, and it was confirmed by the {\it Spitzer}/IRS spectra later obtained for a small sample of MBs \citep{Flagey2011b, Nowak2014}.

Optical and near-IR spectroscopic observations of the central sources in MBs discovered by {\it Spitzer} have successfully revealed several new WR and LBV candidates \citep{Gvaramadze2010, Wachter2010, Wachter2011,Stringfellow2012a}. Most of these observations were performed on MBs with bright central sources detected at 24~$\mu$m. In this paper, we extend the identification of central stars in MBs to those fainter or not detected in the MIPSGAL~24~$\mu$m images, using observations made at the Palomar Observatory with the near-IR Triple Spectrograph. Our sample comprises 14 MBs and 3 comparison sources.

The paper is organized as follows. In section \ref{sec:obs} we detail the observations and data reduction. In section \ref{sec:res} we present our results, starting with the identification of the sources using the most prominent spectral features, and comparing our spectra to those in published libraries. We then characterize the few circumstellar shells that were detected in our spectroscopic observations. We summarize the recent findings on the natures of the MBs collected over the past four years in section \ref{sec:sum}. Our conclusions are listed in section \ref{sec:ccl}.

\begin{table*}[]
  \centering
  \caption{\label{tab:tgt} Summary of the observations}
  \begin{tabular}{c c c c c c c c c c c c c c}
    \hline
    Target & Official nomenclature & 2MASS source & \multicolumn{3}{c}{Magnitudes} & Classification \\
     &  &  & {\it J} & {\it H} & {\it K} &  &  \\
    \hline
    \hline
    MB3157 & MGE012.0365+01.0152 & 18081166-1801004 & 15.4  &  13.8  &  13.3 & K0-K1~III &  \\
    MB3214 & MGE019.6490+00.7742 & 18240393-1126150 & 14.2  &  13.2  &  12.7 & K2-K5~III$^j$ &  \\
    MB3222 & MGE030.1502+00.1239 & 18455526-0225089 & 17.7  &  14.2  &  11.7 & Be/B[e]/LBV$^j$ &  \\
    MB3259 & MGE023.4497+00.0822 & 18334346-0823353 & 14.4  &  12.0  &   9.8 & Be/B[e]/LBV &  \\
    MB3367 & MGE032.4980+00.1616 & 18500440-0018455 & 16.2  &  14.3  &  12.9 & O5-6~V?$^i$ &  \\
    MB3438 & MGE042.0785+00.5085 & 19062457+0822015 & 9.7  &   7.9   &   6.9 & Be/B[e]/LBV$^b,h$ \\
    MB3448 & MGE042.7663+00.8224 & 19063366+0907206 & 15.6  &  14.9  &  13.9 & WC5-6$^e$ \\
    MB3458 & MGE044.5868+00.7929 & 19100426+1043292 & 15.2  &  14.7  &  13.8 & G8-K2~III &  \\
    MB3611 & MGE048.7813-00.8564 & 19240333+1339493 & 10.2  &   9.2  &   8.6 & Be/B[e]/LBV$^c$ \\
    MB3643 & MGE042.9677-01.0182 & 19133273+0827030 & 12.4  &  11.0  &  10.5 & K2-K5~III &  \\
    MB3662 & MGE038.7425-00.6984 & 19043348+0450567 & 13.2  &  12.2  &  11.7 & B/A~III & \\
    MB3671 & MGE037.9741-00.7952 & 19032945+0407201 & 16.1  &  14.6  &  14.0 & G4-G6~III & \\
    MB3701 & MGE032.4004-00.1505 & 18510028-0032309 & 14.1  &  12.1  &  11.0 & B0~I & \\
    MB3834 & MGE019.8601-00.3578 & 18283340-1146441 & 14.4  &  11.4  &   9.7 & Be/B[e]/LBV$^d$ \\
    \hline
    BD+43~3710 & & 20453472+4332271 & 6.6  &  6.1  &  5.9 & B5~I$^a$ &  \\
    WR138a & & 20170811+4107270 & 10.2  &  9.3  &  8.6 & WN9h$^f$  \\
    WR149 & & 21071169+4825361  &  10.6  &  10.1  &  9.6 & WN5sh$^g$   \\
    \hline
  \end{tabular}
  \tablecomments{$^a$BD+43~3710 was wrongly associated by \citet{Kraemer2010} with the nearby carbon star V2040~Cyg first discovered by \citet{Nassau1954}, see section \ref{sec:oba} for details.
$^{b,c,d}$These stars were identified as B[e]/LBV, Be, and B[e]/LBV respectively by \citet{Wachter2011}.
$^e$This star was identified as a [WC] by \citet{Gvaramadze2010}.
$^f$This star was identified as a WN8-9h by \citet{Gvaramadze2009}.
$^g$This star was identified as a WN5s by \citet{Hamann2006}.
$^h$This star was identified as a candidate LBV by \citet{Stringfellow2012a}.
$^i$This nebula was suggested as a PN candidate by \cite{Ingallinera2014}.
$^j$These nebulae are listed as PNe in SIMBAD, as suggested by \citet{Miszalski2008, Urquhart2009}.}
\end{table*}

\section{Observations}
\label{sec:obs}

We acquired the data on August 21st and 22nd, 2010 at the Hale 200 inch telescope of the Palomar Mountain Observatory using the near-IR Triple Spectrograph (TripleSpec). TripleSpec uses a 1\arcsec\ by 30\arcsec\ slit and obtains a {\it JHK}, 1.0 to 2.4~$\mu$m, spectrum at a resolution of 2500-2700, with $\sim$2.7 pixels per resolution element over a 1024x2048 pixel array. During our observations, the seeing was about 1\arcsec\ or better.

The targets were chosen among the MBs that were unidentified at the time of the call for proposals, and that show at least one potential central source in the 2MASS or {\it Spitzer}/IRAC images (see Figure \ref{fig:rgb}). For each of these MBs, we selected the star that appears as the best candidate for the actual central source. We also observed three comparison sources: the WN star WR149 \citep{Hamann2006} and two objects in the Cygnus-X region that resemble some of the MBs: BD+43~3710 and WR138a, identified as a carbon star \citep{Kraemer2010} and a WR star \citep{Gvaramadze2009} respectively. Table \ref{tab:tgt} lists our targets and their near-IR magnitudes. The average {\it K} band magnitude of our sample ($\sim$11.5) was about 2 magnitudes deeper than the samples of \citet{Gvaramadze2010} and \citet{Wachter2010}. Included also in Table \ref{tab:tgt} are the references for the MBs in our sample that have been identified since the time of our observations (MB3438, MB3448, MB3611, and MB3834). MB3214 and MB3222 are listed as PNe in SIMBAD. MB3214 is detected in H$\alpha$ and is listed as a True PN in the MASH-II catalog \citep{Miszalski2008}. MB3222 has been observed in the 6~cm radio survey of candidates massive young stellar objects by \citet{Urquhart2009} who suggested it is a PN.

The length of the slit allowed us to use the usual AB nodding observational strategy in most cases: the source was positioned about 5-10\arcsec\ off center for a first exposure (A) before being moved to the opposite side of the slit for a second exposure (B). An immediate sky subtraction is thus performed while spending 100\% of the time on-source. However, because our sources are in very crowded fields of the Galactic plane, we could not use this strategy for all the targets. In those cases, we spent only 50\% of the time on-source and the remaining 50\% were spent on a selected off-source position. Typical exposure times ranged from 20 to 300 seconds.

\subsection{Data reduction}

\begin{figure*}[t]
  \centering
  \subfigure[]
  {\label{fig:latetype_J}
    \includegraphics[angle=90, width=.45\linewidth]{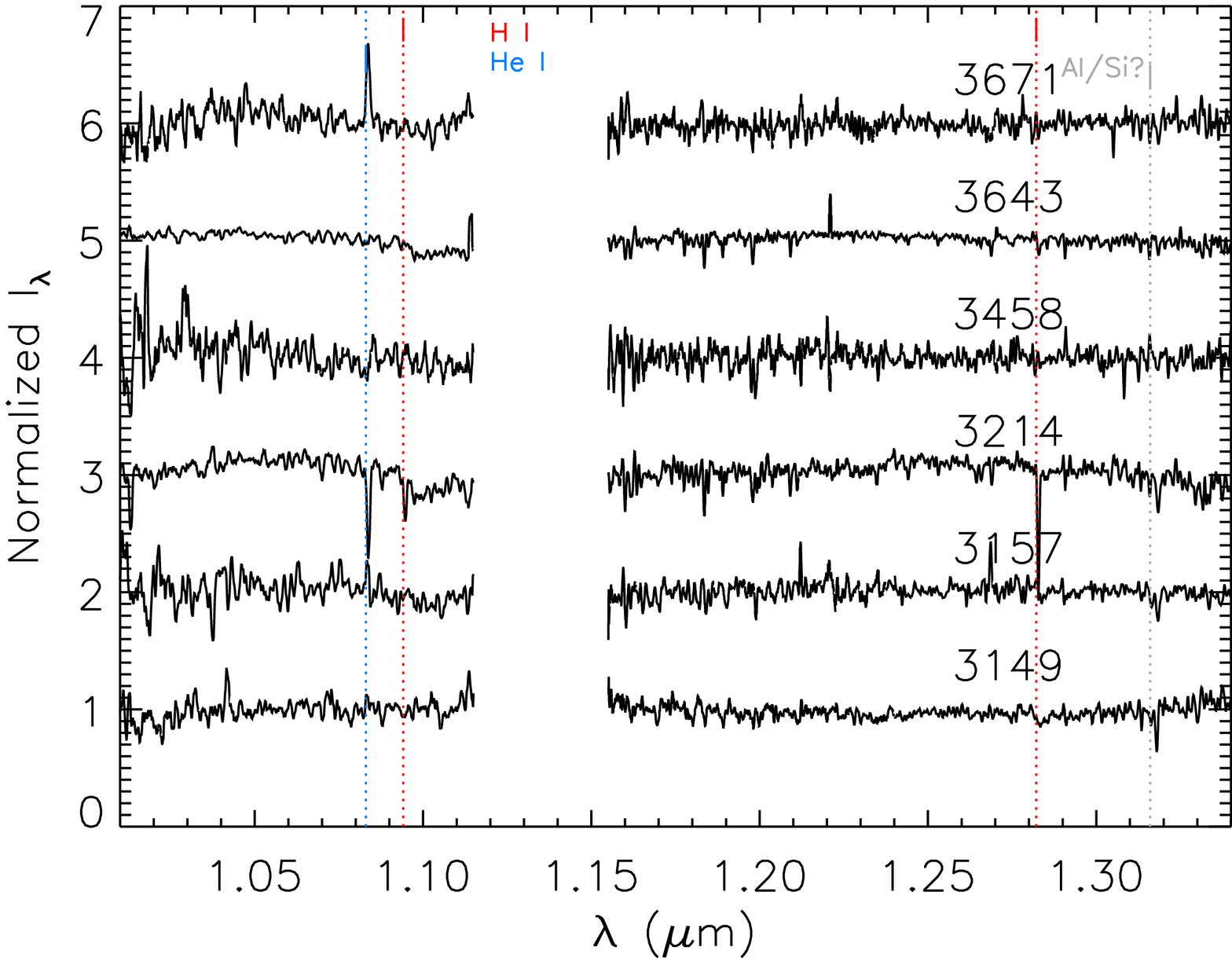}}
  \subfigure[]
  {\label{fig:latetype_H}
    \includegraphics[angle=90, width=.45\linewidth]{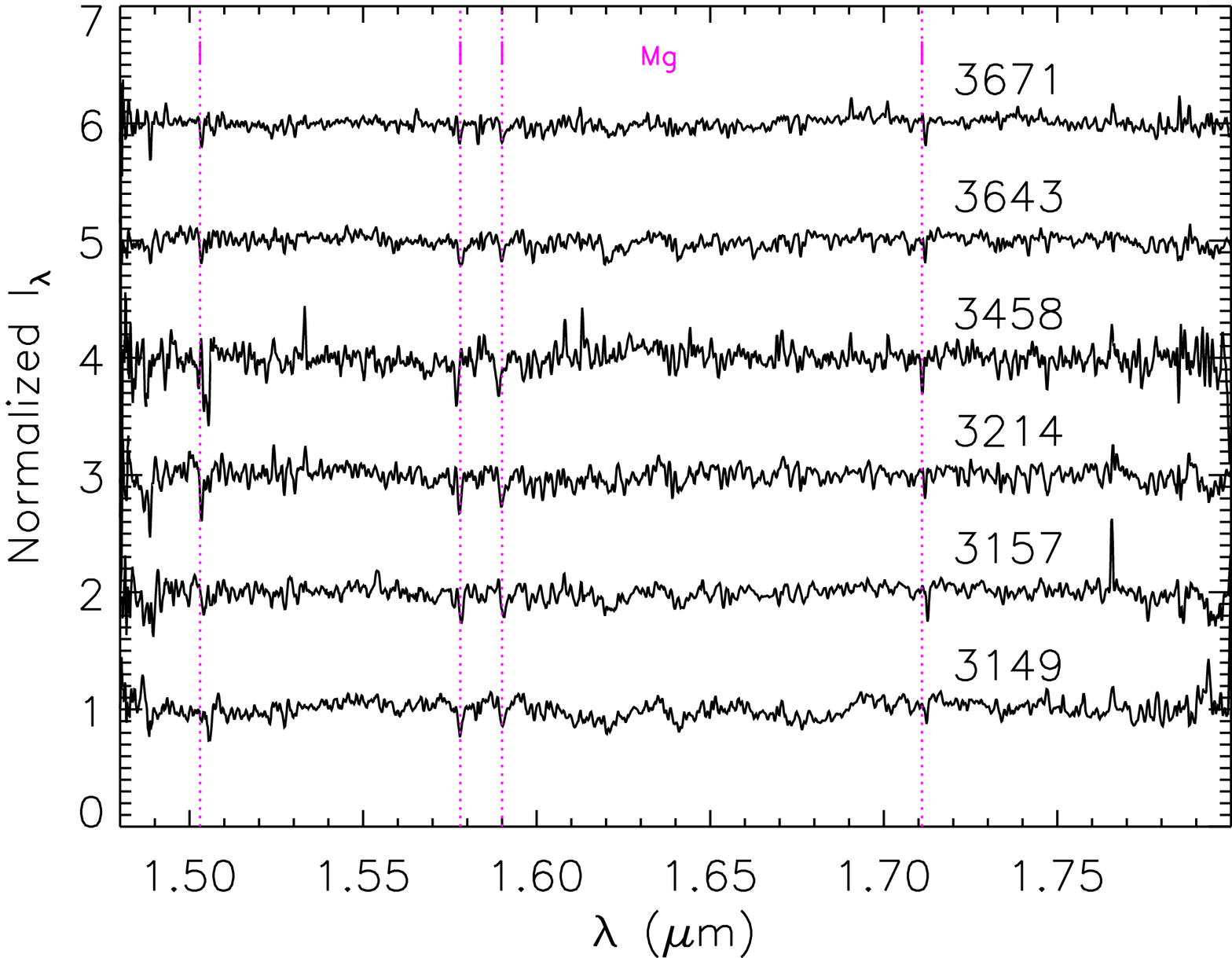}}
  \subfigure[]
  {\label{fig:latetype_K}
    \includegraphics[angle=90, width=.45\linewidth]{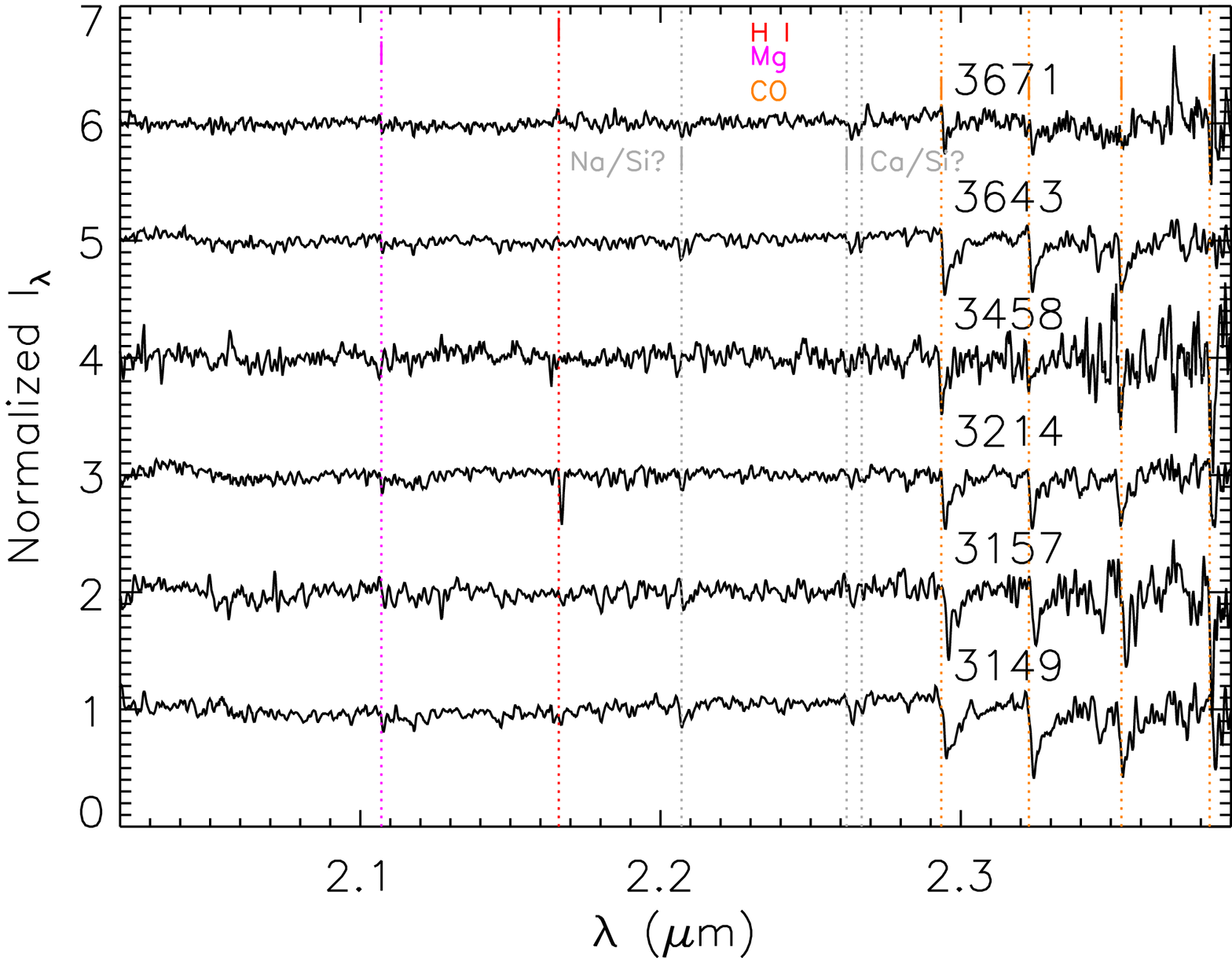}}
  \caption{Near-IR spectra (a: {\it J} band, b: {\it H} band, c: {\it K} band) of the late type stars in our program. The most prominent lines are indicated by vertical dashed lines of different color for different species.}
  \label{fig:latetype_spec}
\end{figure*}

The data reduction was performed with XSpexTool v4.0beta for TripleSpec at Palomar developed by W. Vacca and M. Cushing \citep{Cushing2004, Vacca2004}. This IDL widget package extracts the spectrum from each exposure, taking into account the wavelength calibration performed on the atmospheric lines, the bias and the flat field corrections, and then combines all the exposures together. We did not use XSpexTool to remove the telluric absorption and calibrate the spectra because it assumes the standards are of A0~V type while we observed F and G type stars. Instead we used our own routine and F and G star spectral templates from the IRTF Spectral Library \citep{Rayner2009}. For each target we tried several standards for which the airmass and the time of observation were close to those of the source of interest and chose the final spectrum that shows the best overall quality in terms of signal-to-noise ratio and low atmospheric line residuals. The resulting spectra are shown in Figures \ref{fig:latetype_spec}, \ref{fig:othertype_spec}, \ref{fig:wrtype_spec}, and \ref{fig:lbvtype_spec}, after normalization by a polynomial fit to the continua.

\section{Results}
\label{sec:res}


The 17 targets in this program can roughly be separated into two groups: eight stars with near-IR spectra dominated by absorption lines and nine stars whose spectra show many emission lines. For each of these two groups, two sub-groups can be defined. In the ``absorption''-dominated group, five targets exhibit the CO band-head absorption features at 2.29~$\mu$m and beyond, characteristic of G and later type giants and supergiants, while three stars show hydrogen and helium transitions in absorption, more typical of B and A type stars. In the ``emission''-dominated group, four sources exhibit the \ion{He}{2}~2.189\mic\ line characteristic of O and WR stars, while the five other targets are set apart by the absence of the \ion{He}{2} line and the presence of metal lines in emission, which are more common in LBV candidates. We discuss each sub-group separately. We close this section with the unexpected detection of several shells in our observations.

\subsection{Late type stars: G and K giants}
\label{sec:late}

The near-IR spectra of the five stars that exhibit CO band-head absorption features are shown in Figure \ref{fig:latetype_spec}. The {\it K} band is dominated by the $\Delta v=2$ band-head features at 2.29\mic\ and beyond. Several other absorption lines due to metals can be found (e.g. Mg at 1.58 and 1.59\mic, Si or Al at 1.31\mic, Si or Na at 2.2\mic, Si or Ca at 2.26\mic) although they are significantly weaker and their identification thus remains difficult.

\begin{table*}[]
  \centering
  \caption{\label{tab:co} Equivalent width of the CO 2.29~$\mu$m band-head features and inferred spectral type, extinction, distance, and shell size.}
  \begin{tabular}{c c c c c c c c c c}
    \hline
    Target & $E_{CO}$ ($\AA$) & Spectral type & $A_V$ & d (kpc) & 24~$\mu$m radius & 24~$\mu$m size (pc) \\
    \hline
    \hline
    MB3157 & 10.1$^a$ / 16.8$^b$ & G6-G7~I$^a$ / G8-G9~I$^b$ & 9.0$\pm$0.1 & 120$\pm$3 & 35\arcsec & 20 \\
     &  & \bf{K0-K1~III$^a$ / K0-K1~III$^b$} & \bf{8.3$\pm$0.1} & \bf{6.8$\pm$0.5} &  & \bf{1.2} \\

    MB3214 & 11.9$^a$ / 24.8$^b$ & G7-G8~I$^a$ / K0-K1~I$^b$ & 5.4$\pm$0.3 & 112$\pm$5 & 20\arcsec & 10 \\
     &  & \bf{K2-K3~III$^a$ / K4-K5~III$^b$} & \bf{3.7$\pm$0.1} & \bf{11.8$\pm$4.1} &  & \bf{1.1} \\

    MB3458 & 7.2$^a$ / 19.3$^b$ & G4-G5~I$^a$ / G9-K0~I$^b$ & 5.0$\pm$0.3 & 183$\pm$10 & 25\arcsec & 22 \\
     &  & \bf{G8-G9~III$^a$ / K1-K2~III$^b$} & \bf{4.3$\pm$0.4} & \bf{10.4$\pm$2.1} &  & \bf{1.3} \\

    MB3643 & 14.3$^a$ / 21.3$^b$ & G9-K0~I$^a$ / G9-K0~I$^b$ & 7.7$\pm$0.1 & 36.3$\pm$0.5 & 45\arcsec & 7.7 \\
     &  & \bf{K4-K5~III$^a$ / K2-K3~III$^b$} & \bf{6.5$\pm$1.1} & \bf{3.4$\pm$1.8} &  & \bf{0.7} \\

    MB3671 & 4.7$^a$ / 6.9$^b$ & G2-G3~I$^a$ / G6-G7~I$^b$ & 9.3$\pm$0.2 & 157$\pm$7 & 20\arcsec & 15 \\
     &  & \bf{G5-G6~III$^a$ / G4-G5~III$^b$} & \bf{8.8$\pm$0.1} & \bf{7.1$\pm$0.3} &  & \bf{0.7} \\
    \hline
  \end{tabular}
  \tablecomments{$^a$ values derived using the method of \citet{Davies2007}, $^b$ values inferred using the method of \citet{Figer2006}. For each of the 5 late type stars, the first line gives the parameters for a supergiant while the second line gives those for a giant. In {\bf bold} is our suggested identification.}
\end{table*}

\citet{Figer2006} and \citet{Davies2007} have shown that there is an anticorrelation between $E_{CO}$, the equivalent width of the 2.29\mic\ feature, and the star's temperature. We measure $E_{CO}$ the same way it was done by \citet{Figer2006} and \citet{Davies2007}. The two estimates of the equivalent width are reported in Table \ref{tab:co} for the five late-type stars in our sample. The classifications then inferred from the work of \citet{Figer2006} and \citet{Davies2007} usually agree within a few sub-types at most. Assuming the near-IR colors and absolute magnitudes for red giants and supergiants from Allen's Astrophysical Quantities \citep{Cox2000}, assuming the interstellar extinction curve from \citet{Cardelli1989}, and using the 2MASS {\it J} and {\it K} band magnitudes, we derive the visual extinction towards each source and its distance. We rule out the supergiant interpretation for each of them, based on the average extinction along the line of sight. On average, we expect about 2 mag/kpc of visual extinction using $A_V/N_H=0.53\times 10^{-21}~\rm{cm^2}$ and an average interstellar gas density of 1~cm$^{-3}$ along the line of sight \citep[also see][]{Whittet2003}. In the supergiant scenario, the five sources would be along lines of sight with at most 0.2~mag/kpc. The derived distances also rule out the supergiant interpretations for all targets, under the hypothesis that these are Galactic objects. In the supergiant scenario, the five sources would instead be located between 36 and 183~kpc from the Sun and have sizes in the range 8-22~pc.

For the preferred giant interpretation, we derive the size of the 24~$\mu$m shell for each late-type star. We find that the five shells are between 0.7 and 1.3~pc in radius. These sizes are larger by at least a factor of two than those derived by \citet{Young1993} from IRAS~60~$\mu$m observations of red giant stars and young PNe (less than 0.3~pc mostly). This difference might be due to a distance selection effect, because IRAS could only resolve nearby ($\lesssim1$~kpc) objects. They are smaller than the sizes found by \citet{Stencel1989} in their analysis of supergiant stars (between 1 and 3~pc), which also used IRAS~60~$\mu$m observations. \citet{Wachter2010} found 12 late-type stars in their sample of objects similar to ours, and suggested that all but one are supergiants. The shell sizes they derived range from 0.4~pc for the only suggested giant, to 0.5-3~pc for the supergiants. These sizes are consistent with the sizes we inferred for our sample.

\begin{figure*}[t]
  \centering
  \subfigure[]
  {\label{fig:othertype_J}
    \includegraphics[angle=90,width=.45\linewidth]{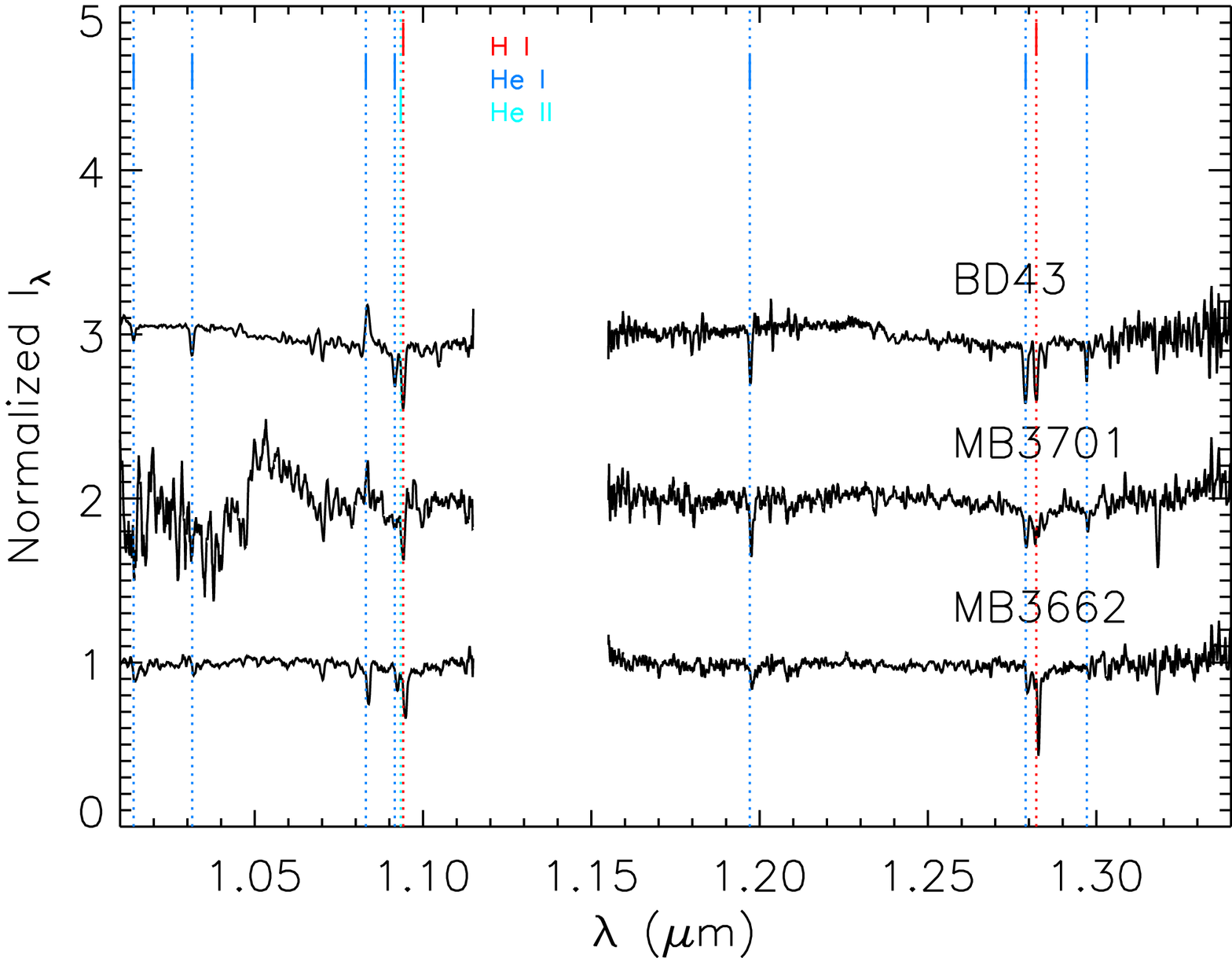}}
  \subfigure[]
  {\label{fig:othertype_H}
    \includegraphics[angle=90,width=.45\linewidth]{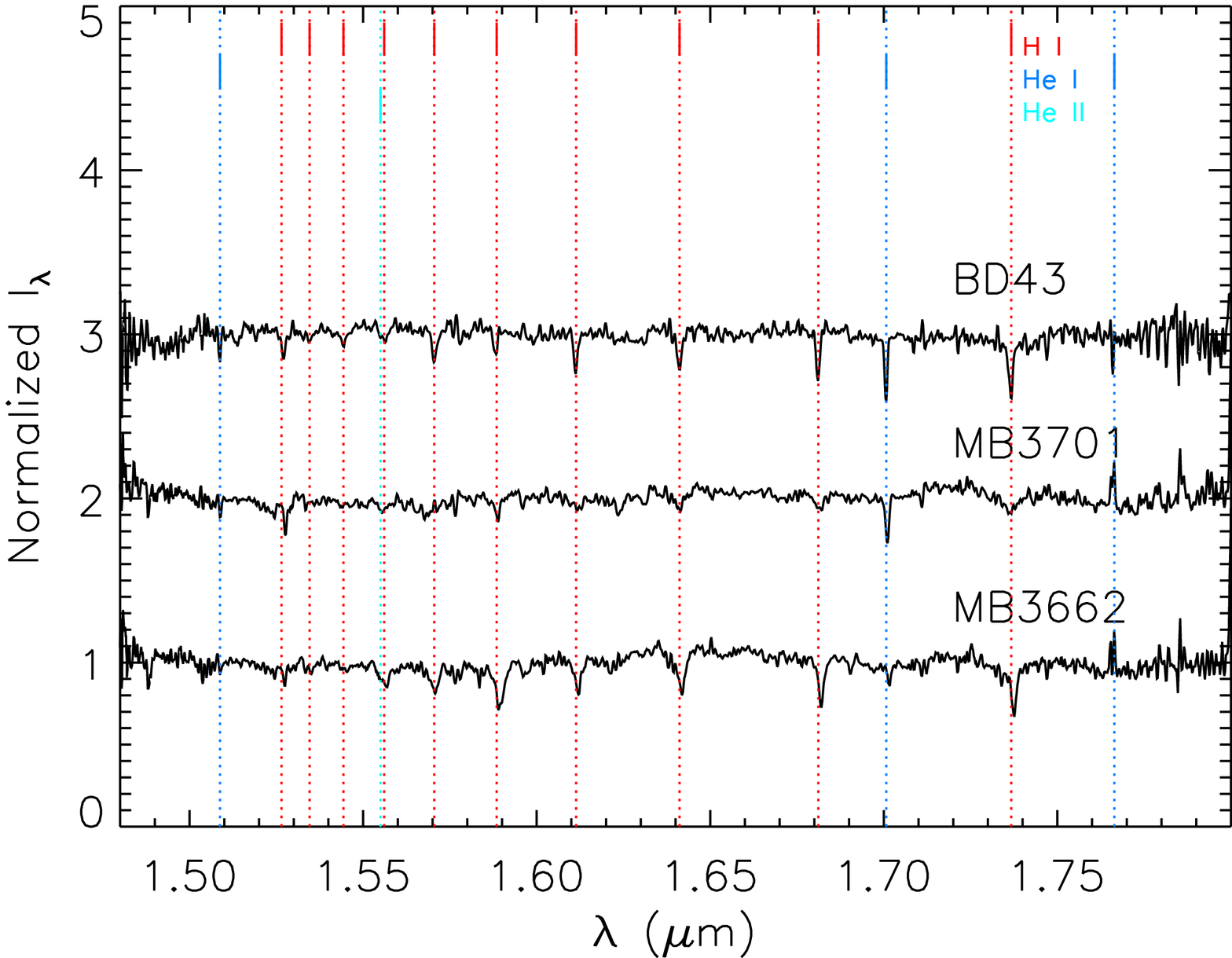}}
  \subfigure[]
  {\label{fig:othertype_K}
    \includegraphics[angle=90,width=.45\linewidth]{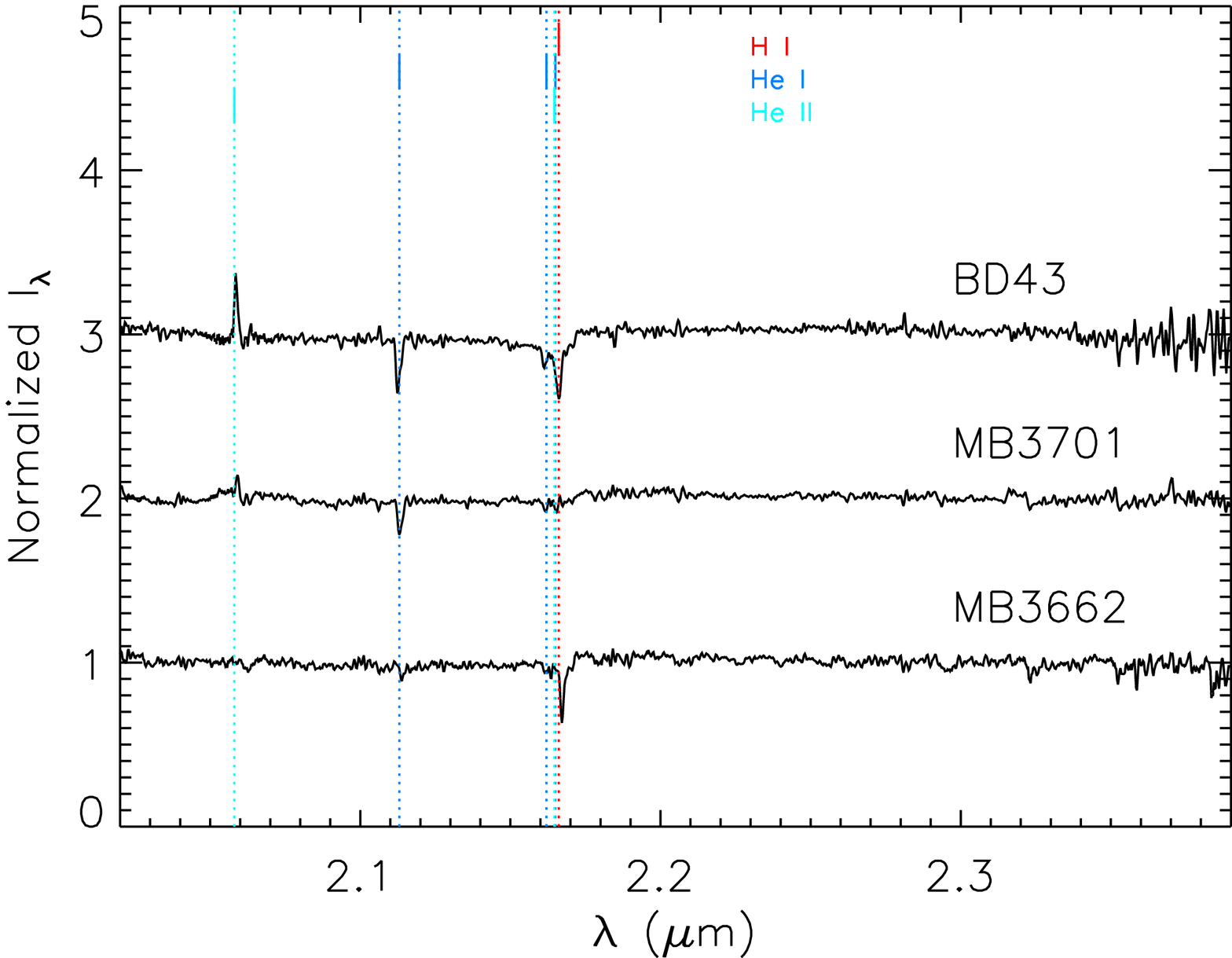}}
  \caption{Near-IR spectra (a: {\it J} band, b: {\it H} band, c: {\it K} band) of the OBA type stars in our program. The most prominent lines are indicated by vertical dashed lines of different color for different species.}
  \label{fig:othertype_spec}
\end{figure*}

\subsection{Early type stars: B and A stars}
\label{sec:oba}

Three targets in our sample have spectra dominated by hydrogen and/or helium transitions in absorption, which are characteristics of early type (B and A) stars. These targets are BD+43~3710, MB3662, and MB3701 (see Figure \ref{fig:othertype_spec}).

{\it BD+43~3710} - This source has been associated with a carbon star of spectral type R: by \citet{Kraemer2010} based on the initial discovery by \citet{Nassau1954}. However, the table where the actual carbon star was found in the latter paper ``gives a reference BD star near the carbon star'', implying that the carbon star is not BD+43~3710. \citet{Nassau1954} wrote that the carbon star was found $0.6$~mm to the East and $1.2$~mm to the South of the BD star, ``in the scale of the BD chart from the reference star to the carbon star''. The plate scale of the Bonner Durchmusterung is 3\arcmin\ per mm, which locates the carbon star about 1.8\arcmin\ to the East and 3.6\arcmin\ to the South of BD+43~3710. In the 2MASS {\it JHK} images near BD+43~3710, a bright and red star ({\it J}=5.55, {\it H}=4.046, and {\it K}=3.26) is visible at about 1.8\arcmin\ to the East and 3.2\arcmin\ to the South. This star is 2MASS 20454509+4329181, also known as V2040~Cyg, and is referenced as a semi-regular pulsating carbon star in SIMBAD \citep{Aaronson1990}. A spectral type N5 was found for V2040~Cyg by \citet{Eglitis2003}. Its R and J magnitudes of 10.8 and 6.6, respectively, seem like a good match to the infrared magnitude $m_i$ of about 9 reported by \citet{Nassau1954}. We therefore suggest that the carbon star of \citet{Nassau1954} is V2040~Cyg.

The near-IR spectrum of a carbon star usually shows broad, molecular features of CN and C$_2$ over the entire near-IR range, as well as the CO band-head absorption features at 2.29\mic\ and beyond \citep[e.g.][]{Rayner2009}. The near-IR spectrum of BD+43~3710 however shows strong absorption in the hydrogen transitions of the Brackett and Paschen series, and in the \ion{He}{1} transitions. The strength of the H absorption lines, combined with the detection of \ion{He}{1} lines in absorption indicates a good match to B type stars \citep[see e.g.][]{Lancon1992, Hanson2005}. The best match is the B5 supergiant HR2827 \citep{Meyer1998}. We favor this luminosity class because of the narrow absorption lines. Giants and dwarfs usually have significantly broader absorption features than supergiants, because of their higher gravity.

To test the validity of our interpretation of BD+43~3710, we use the intrinsic colors for early supergiants \citep{Allen1983} and estimate its distance and the extinction along the line of sight. We use the B and V magnitudes referenced in SIMBAD, as the near-IR magnitudes may be contaminated by hot dust emission. Assuming a B5~I spectral classification, we find that BD+43~3710 is at about 2.5~kpc, just behind the Cygnus-X region, and that the total visual extinction along the line of sight is 3~mag. At 2.5~kpc, the 2.7\arcmin\ by 0.9\arcmin\ mid-IR shell of BD+43~3710 would have a physical size of 2.0 by 0.7~pc.

{\it MB3701 and MB3662} - The spectra of the stars at the centers of these MBs are slightly different from that of BD+43~3710. The former shows weaker absorption in the hydrogen transitions but stronger in \ion{He}{1}, while the latter shows stronger hydrogen but weaker \ion{He}{1} absorption lines. This suggests an earlier type for the star at the center of MB3701 than for BD+43~3710, and a later type for that in MB3662. As for the BD star, we favor the supergiant luminosity class based on the width of the absorption features for the central star of MB3701, although the limited number of these features prevents us from reaching a clear conclusion. For example, the spectrum of HR1903 \citep[B0~I,][]{Meyer1998, Wallace2000} is a good match to that of the central source in MB3701. For the star in MB3662, the widths of the hydrogen lines in H band are slightly larger than those of the BD star, which could suggest a higher luminosity class. In terms of the absorption features relative intensities, the spectra of HR3975 and HR2827 \citep[A0~I and B5~I respectively,][]{Meyer1998} are both good matches to that of the central star in MB3662, especially because the He line at 1.7\mic~is still present. However, HR1552 and HR5291 (B2~III and A0~III respectively) seem to be better suited in terms of line widths.

Unlike BD+43~3710, the star at the center of MB3662 is faint in the optical ($B\sim21$~mag, $V=18.6$~mag). Assuming $A_B/A_R=1.79$, we find that as an A0 or B5 supergiant, this source would be too far away ($\gtrsim$16~kpc) and with too little extinction along the line of sight ($A_V\sim$8~mag). We therefore favor a higher luminosity class (e.g. B5~III) for this star. We use the intrinsic colors published by \citet{Pickles1998} and find that MB3662 would then be at 3.7~kpc with about 8.3~mag of extinction along the line of sight. At that distance, MB3662 would be about 0.3~pc in radius. For the star at the center of MB3701, only near-IR magnitudes are available. We infer from them that MB3701 would be at about 14~kpc with a visual extinction of about 15~mag, assuming a B0~I spectral classification. The average extinction towards MB3701 would be in fair agreement with the average Galactic value, and the physical radius of MB3701 would be about 1.7~pc. However, an IR excess can contribute to the central source fluxes and a slightly different spectral type or luminosity class remains possible for MB3701.

\subsection{Emission line stars}
\label{sec:emi}
Among the nine spectra that are dominated by emission lines, four exhibit many unresolved emission lines while the five others are dominated by broad lines. Broad lines are an indication of strong stellar winds, which can be found in WR stars, where typical velocities are above 1000~km/s \citep{Crowther2007}. LBV stars are also known for their stellar winds with velocities of a few 100~km/s. Instead of using the presence of broad lines, we use the presence of the \ion{He}{2} lines at 1.012 and 2.189\mic\ as a strong indication that a given source is an O or WR star, following the studies of \citet{Morris1996}, \citet{Figer1997}, \citet{Crowther2006}, and \citet{Wachter2010}.

\begin{figure*}[t]
  \centering
  \subfigure[]
  {\label{fig:wrtype_J}
    \includegraphics[angle=90,width=.45\linewidth]{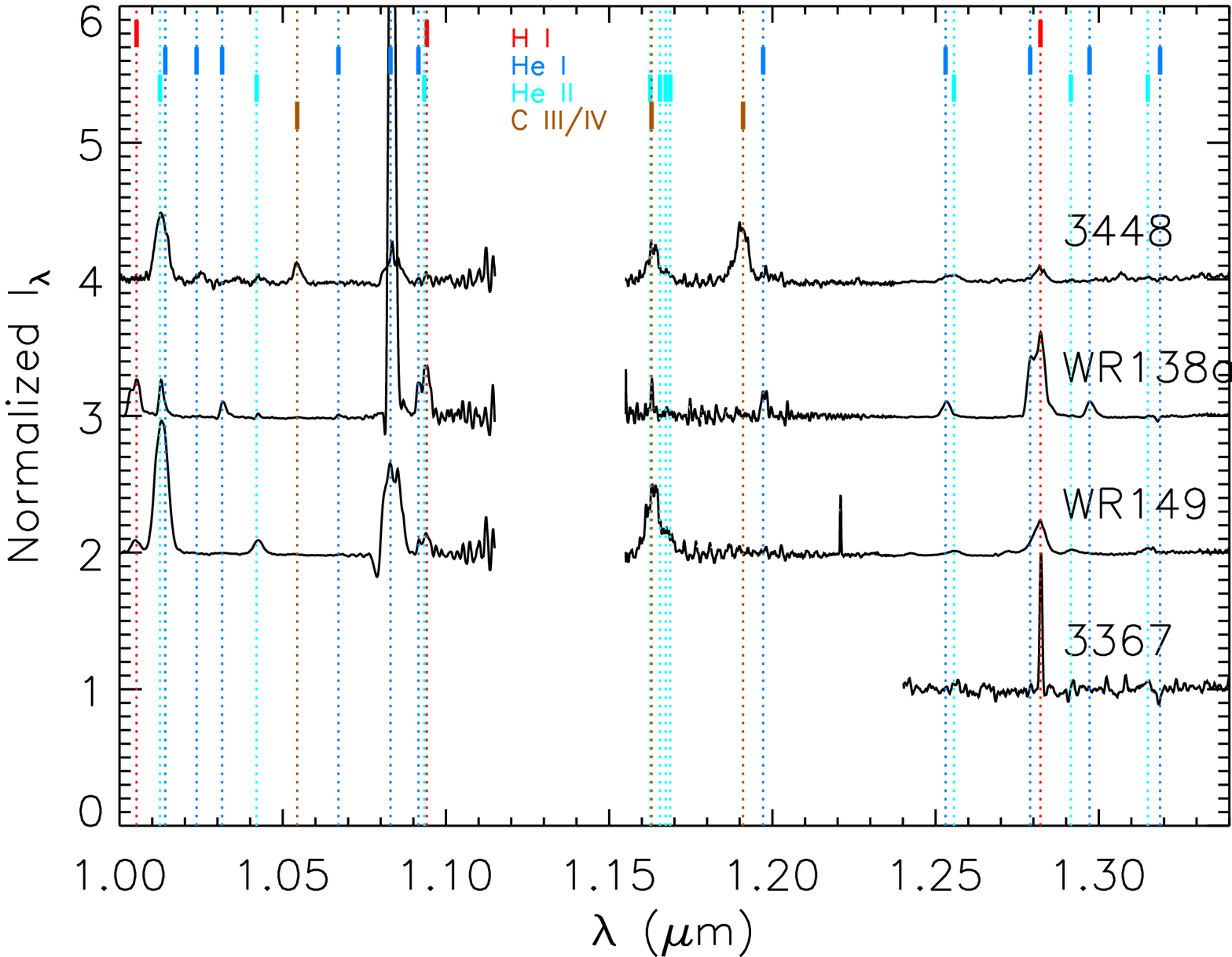}}
  \subfigure[]
  {\label{fig:wrtype_H}
    \includegraphics[angle=90,width=.45\linewidth]{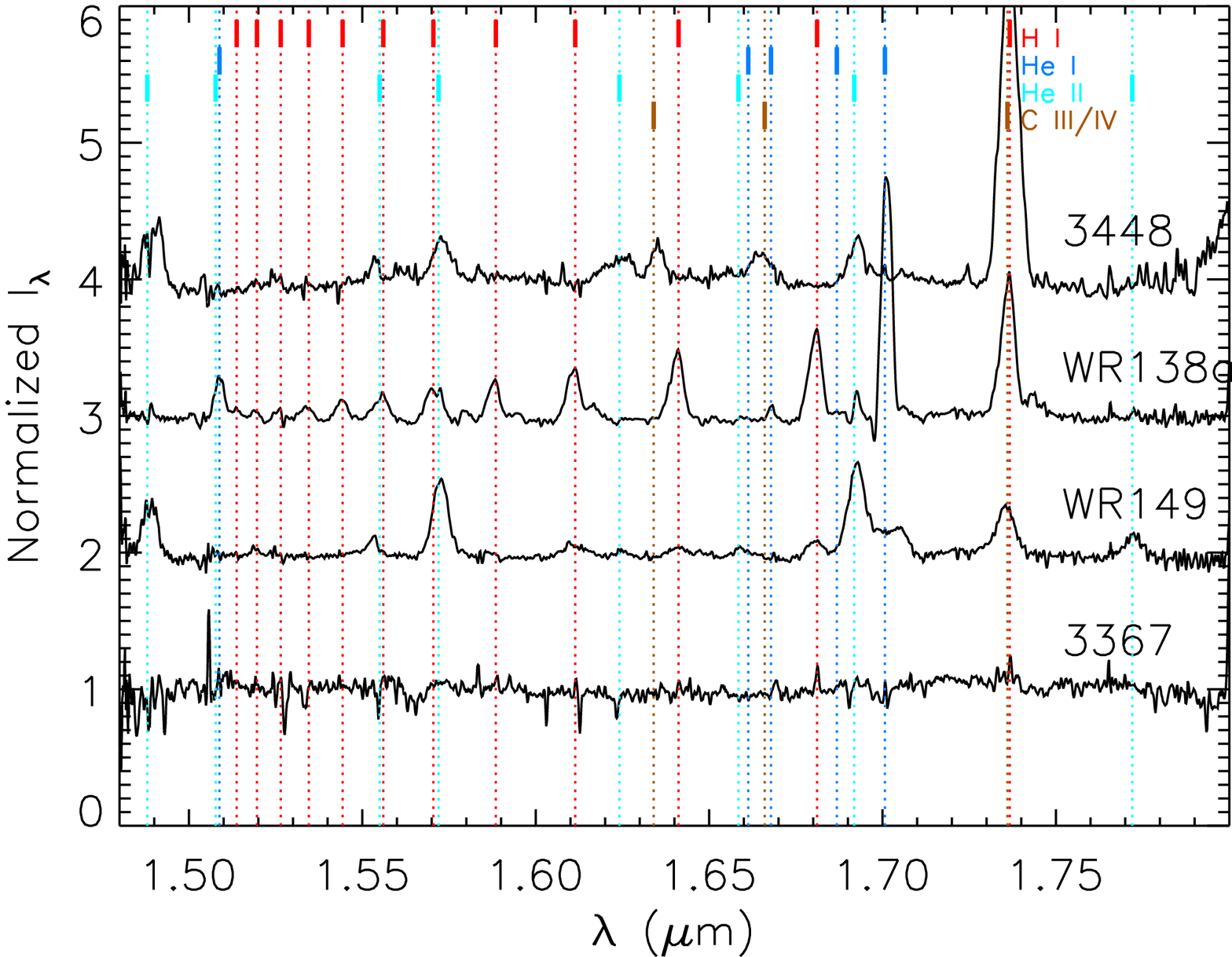}}
  \subfigure[]
  {\label{fig:wrtype_K}
    \includegraphics[angle=90,width=.45\linewidth]{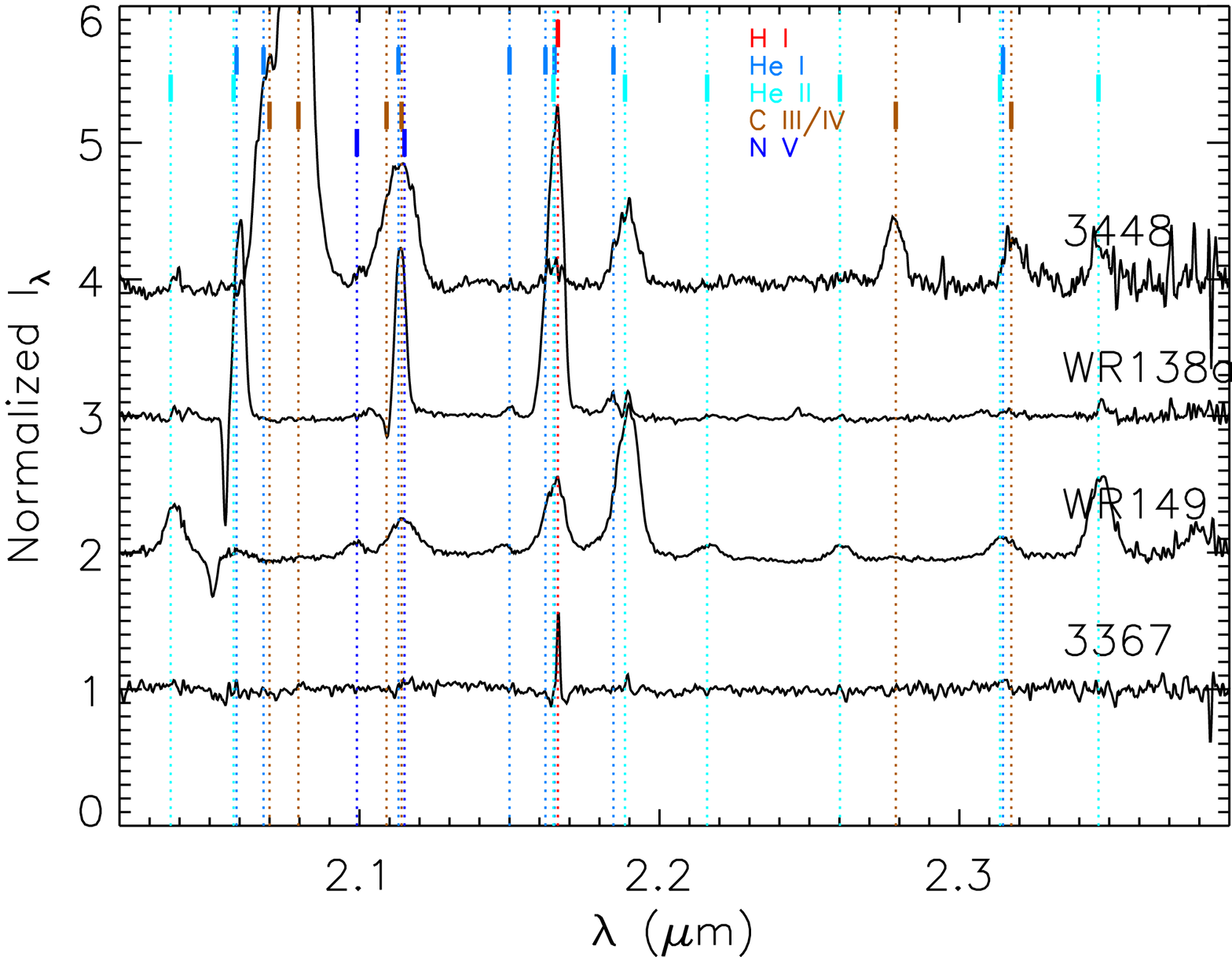}}
  \caption{Near-IR spectra (a: {\it J} band, b: {\it H} band, c: {\it K} band) of the emission line stars that we suggest as O and WR candidates. For the star at the center of MB3367, the short wavelengths end of the {\it J} band spectra are mostly useless and therefore not shown. The most prominent lines are indicated by vertical dashed lines of different color for different species.}
  \label{fig:wrtype_spec}
\end{figure*}

\begin{table*}[]
  \centering
  \caption{\label{tab:early_det} Line detection used for the identification of the early type stars}
  \begin{tabular}{c c c c c c c c c c c c c c}
    \hline
    Target & \ion{He}{2} & \ion{He}{2} & \ion{Mg}{2} & \ion{Na}{2} & \ion{Fe}{2} & \ion{C}{3}/\ion{C}{4} \\
    Target & 1.012\mic & 2.189\mic & 2.138/2.144\mic & 2.206/2.209\mic & 1.69/2.09\mic & 1.20/2.08\mic\\
    \hline
    \hline
    MB3222 & &  & \checkmark  &  & \\
    MB3259 & &  & \checkmark  & ?  & \checkmark  & \\
    MB3438 & &  & \checkmark  & \checkmark  & \checkmark  &\\
    MB3611 & &  & \checkmark  & \checkmark  & \checkmark  & \\
    MB3834 & &  & \checkmark  & ?  & ? & \\
    \hline
    MB3367 & ?  & \checkmark  &  &  &\\
    MB3448 & \checkmark  & \checkmark  &  &  & & \checkmark\\
    WR138a & \checkmark  & \checkmark  &  &   \\
    WR149 & \checkmark  & \checkmark  &   &     \\
    \hline
  \end{tabular}
  \tablecomments{A $\checkmark$ indicates a clear detection while a ? indicates a questionable detection.}
\end{table*}

Four targets (MB3448, MB3367, WR149, and WR138a) show either or both of the \ion{He}{2} lines in their spectra (see Table \ref{tab:early_det}). WR149 and WR138a are both known WR stars with mid and late WN type respectively. The five sources that do not show the \ion{He}{2} lines (MB3222, MB3259, MB3438, MB3611, and MB3834) have also in common that their spectra exhibit many emission lines of metal species: Fe lines at e.g. 2.09 and 1.69\mic, the Mg doublet at 2.138-2.144\mic, and the Na doublet at 2.206-2.209\mic. Such lines have been observed preferentially, but not exclusively, in LBV, B[e], and Be stars \citep[e.g.][]{Morris1996}. MB3438, MB3611, and MB3834 have been suggested as B[e]/LBV, Be, and B[e]/LBV, respectively by \citet{Wachter2011}, and \citet{Stringfellow2012a}. A clear distinction thus seems to emerge between O and WR and LBV candidates, based on the presence of the \ion{He}{2} lines at 1.012 or 2.189\mic, or the presence of emission lines of Fe, Mg or Na.

\begin{table*}[]
  \centering
  \caption{\label{tab:early_eqw} Line equivalent widths used for the determination of the Wolf-Rayet stars subtypes}
  \begin{tabular}{c c c c c c c c c c c c c c}
    \hline
    Target & \ion{He}{2} & \ion{He}{1} & \ion{He}{2} & Br$\gamma$ & \ion{N}{5} & \ion{N}{3}/\ion{He}{1} & \ion{C}{4} & \ion{C}{3} & \ion{C}{4} & \ion{C}{3}\\
    Target & 1.012\mic & 1.083\mic & 2.189\mic & & 2.099\mic & 2.115\mic & 1.191\mic & 1.198\mic & 2.076\mic & 2.110\mic\\
    \hline
    \hline
    MB3367 & & 70 & 0.9 & 5 \\
    MB3448 & & & & & & & 140 & 20 & 410 & 170 \\
    \hline
    WR138a & 13 & 380 & 3 & 94 \\
    WR149 & 170 & 188 & 68 & 33 & 6 & 23 \\
    \hline
  \end{tabular}
  \tablecomments{Units are angstroms.}
\end{table*}

\subsubsection{O and Wolf-Rayet stars}
\label{sec:wr}

The spectra of the four O and WR stars are shown in Figure \ref{fig:wrtype_spec}. The four spectra are quite different from each other. We use the classification diagnostics of \citet{Crowther2006}, based on line equivalent width ratios, to derive the subtype of each WR star. The equivalent widths of the diagnostic lines are indicated in Table \ref{tab:early_eqw}.

{\it WR149} - WR149 is a WN5-s star, where the ``s'' indicates strong lines \citep[][based on optical observations]{Hamann2006}. The near-IR spectrum of WR149 shows strong helium lines with P Cygni profiles, and strong hydrogen lines. Several N lines, including the \ion{N}{5} line at 2.099\mic, which is a clear indicator of early type WN stars, are also detected. The three near-IR diagnostics of \citet{Crowther2006} for WN stars indicate a WN5-6 subtype, in very good agreement with that inferred from the optical observations. Due to the presence of hydrogen lines in emission, we suggest that this star be identified as WN5h. The P Cygni profiles of the \ion{He}{1} lines allow us to estimate the speed of the wind in WR149, using a method similar to that presented by \citet{Voors2000b}. We find a terminal velocity of about 1700~km/s, in good agreement with the 1300~km/s found by \citet{Hamann2006} and with the typical velocities in WN5 stars \citep[1500~km/s,][]{Crowther2007}.

{\it WR138a} - This source is a WN8-9h star \citep[][based on optical observations]{Gvaramadze2009}. The near-IR spectrum of WR138a exhibits strong helium lines with P Cygni profiles, and strong hydrogen lines. However, unlike that of WR149, it does not show nitrogen lines, although the \ion{N}{3}~2.115\mic\ line could be blended with the \ion{He}{1} line at 2.11\mic. The diagnostics of \citet{Crowther2006} indicate a WN9 subtype. A WN8 subtype would have significantly larger \ion{He}{2} 1.012\mic\ to \ion{He}{1} 1.083\mic, and \ion{He}{2} 2.189\mic\ to Br$\gamma$ equivalent width ratios. We suggest that this star be identified as a WN9h to take into account the emission lines of hydrogen. The P Cygni profiles of the \ion{He}{1} lines allow us to estimate the speed of the stellar wind in WR138a, using the same method as for WR149. We find a velocity of about 650~km/s, in excellent agreement with the 700~km/s found by \citet{Gvaramadze2009} and with the typical velocities in WN9 stars \citep[700~km/s,][]{Crowther2007}.

{\it MB3448} - The spectrum of the star at the center of MB3448 is dominated by strong C emission lines with FWHMs of about 1000-1500~km/s, which are characteristic of late type WC stars \citep[1200~km/s for WC9, 1700~km/s for WC8,][]{Crowther2007}. The classification diagnostics of \citet{Crowther2006} for WC stars indicate either a WC5-6 or a WC8 subtype. Because the lines are broad, features are blended and the uncertainties in some equivalent widths may be large. The typical wind velocities for WC5-6 stars \citep[2200~km/s,][]{Crowther2007} are significantly larger than those we measure but the \ion{C}{4}~1.191 to \ion{C}{3}~1.198\mic\ line ratio clearly suggests an earlier type than WC8. We thus favor a narrow-line WC5-6 interpretation. Using intrinsic colors and magnitudes for WC5-6 stars in the optical from \citet{vanderHucht2001} and \citet{Crowther2007}, assuming $m_V\sim18.6$~mag for the star at the center of MB3448 ($B=19.9$ and $R=17.4$), and that $A_B=1.31\times A_V$, we find that MB3448 would be at a distance of about 25~kpc and suffer only about 5~mag of visual extinction. At that distance, MB3448 would be about 1.5~pc in radius. We note however that the intrinsic mangitude of WC6 stars could be significantly lower \citep[$M_v\sim-4.3$ in][]{Sander2012} and the distance even larger. \citet{Kanarek2014x} recently listed this star as a WC6, with an uncertainty of up to two subtypes. Assuming it is a massive WR star, they derive a distance of almost 30~kpc, in good agreement with our estimates. However, the average extinction along the line of sight would be significantly smaller than that expected in the interstellar medium. \citet{Gvaramadze2010} identified the star at the center of MB3448 as a [WC] but did not show a spectrum. They favor the low-mass WR interpretation because MB3448 has an optical counterpart, which suggests a distance significantly smaller than those derived in the high-mass WR case. There are no tabulated values for the intrinsic colors of [WR] stars. \citet{Parker2003} found $M_V$ between -1 and 0 for a few [WC] stars. \citet{Leuenhagen1996} gave parameters for five [WC] stars, from which values of $M_V$ between -2.7 and +1 are derived. Assuming the colors of a [WC5-6] are the same as those of a WC5-6, and $M_V\sim-0.5$, MB3448 would be at a distance of $\sim$6~kpc. The shell would then be about 0.5~pc in radius.

{\it MB3367} - The spectrum of the central source in MB3367 is dominated by a few, unresolved emission lines (Pa$\beta$, Br$\gamma$, \ion{He}{2} line at 2.189\mic) and several absorption features (e.g. Br$\gamma$, \ion{He}{1} at 1.70\mic). Possible emission is detected at 2.11-2.12\mic\ (\ion{He}{1} or \ion{N}{5}). The overall signal-to-noise ratio of the spectrum is among the lowest of our sample, due in part to large variations of airmass during the observations. The \ion{He}{2} 2.189\mic\ to Br$\gamma$ equivalent width ratio indicates a WN8 subtype \citep{Crowther2006}, but one would expect a strong stellar wind \citep[$\sim$1000~km/s,][]{Crowther2007} while the lines are unresolved in our spectra ($\le130$~km/s). The shape of the spectrum around Br$\gamma$ evokes that of an O star with nebular emission. \citet{Furness2010} obtained {\it H} and {\it K}-band spectra of O stars in W31: their \#5 in particular, an O5-6~V star, is a very good match to the central star in MB3367. Using near-IR colors for O5 dwarfs \citep{Allen1983}, we find that MB3367 is behind $\sim$23~mag of visual extinction and at a distance of about 11~kpc. As a supergiant, it would be at about 16~kpc, although the apparent widths of the absorption features in the spectrum suggest a lower luminosity class. At a distance of 11~kpc, MB3367 would have a radius of about 0.5~pc. \citet{Ingallinera2014} suggested that MB3367 is a PN candidate based on 6 and 20~cm observations, but could not entirely rule out the thermal emission hypothesis. While we favor the O5-6~V nature for the star at the center of MB3367, we cannot completely rule out other interpretations.

\begin{figure*}[t]
  \centering
  \subfigure[]
  {\label{fig:earlytype_J}
    \includegraphics[angle=90,width=.45\linewidth]{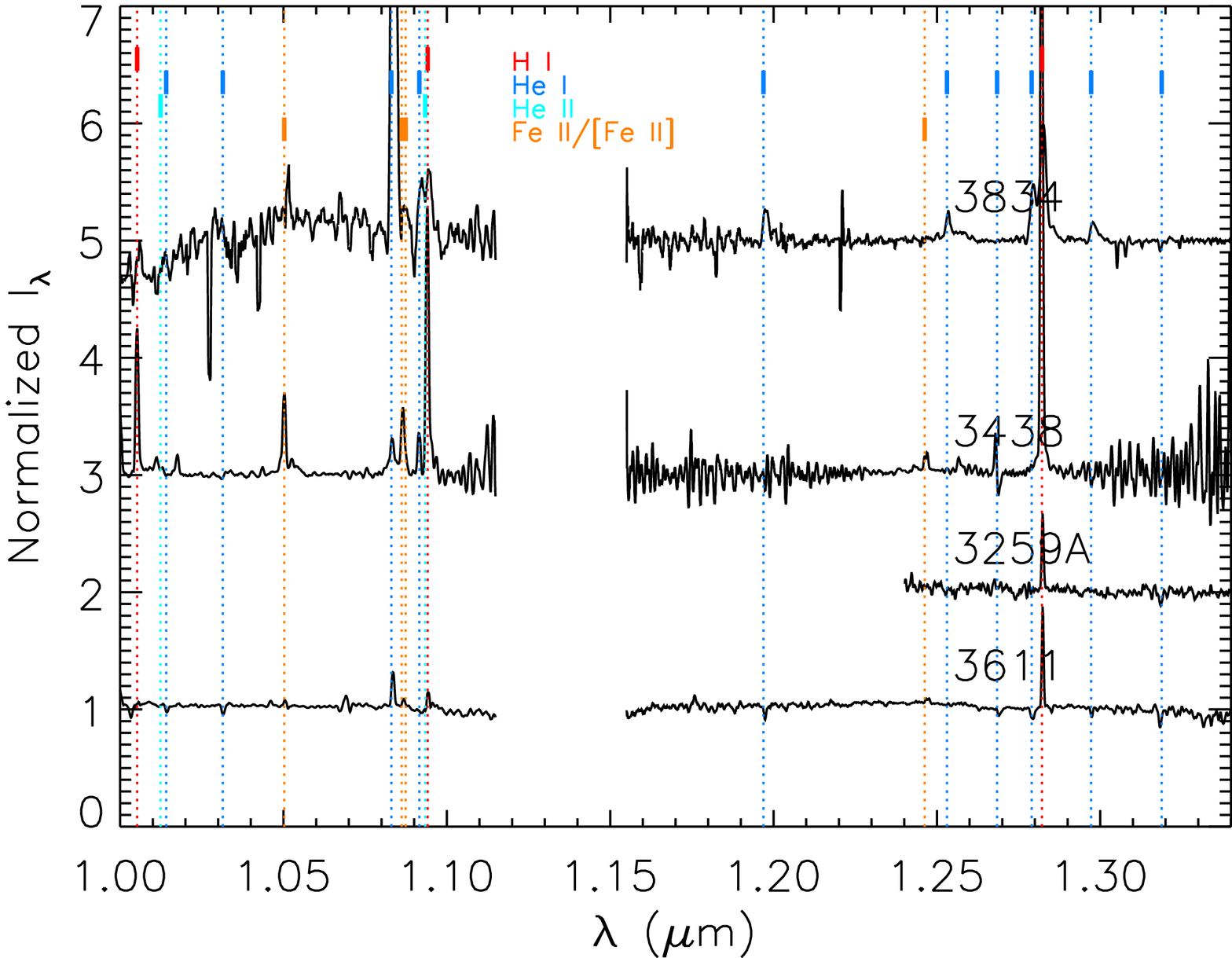}}
  \subfigure[]
  {\label{fig:earlytype_H}
    \includegraphics[angle=90,width=.45\linewidth]{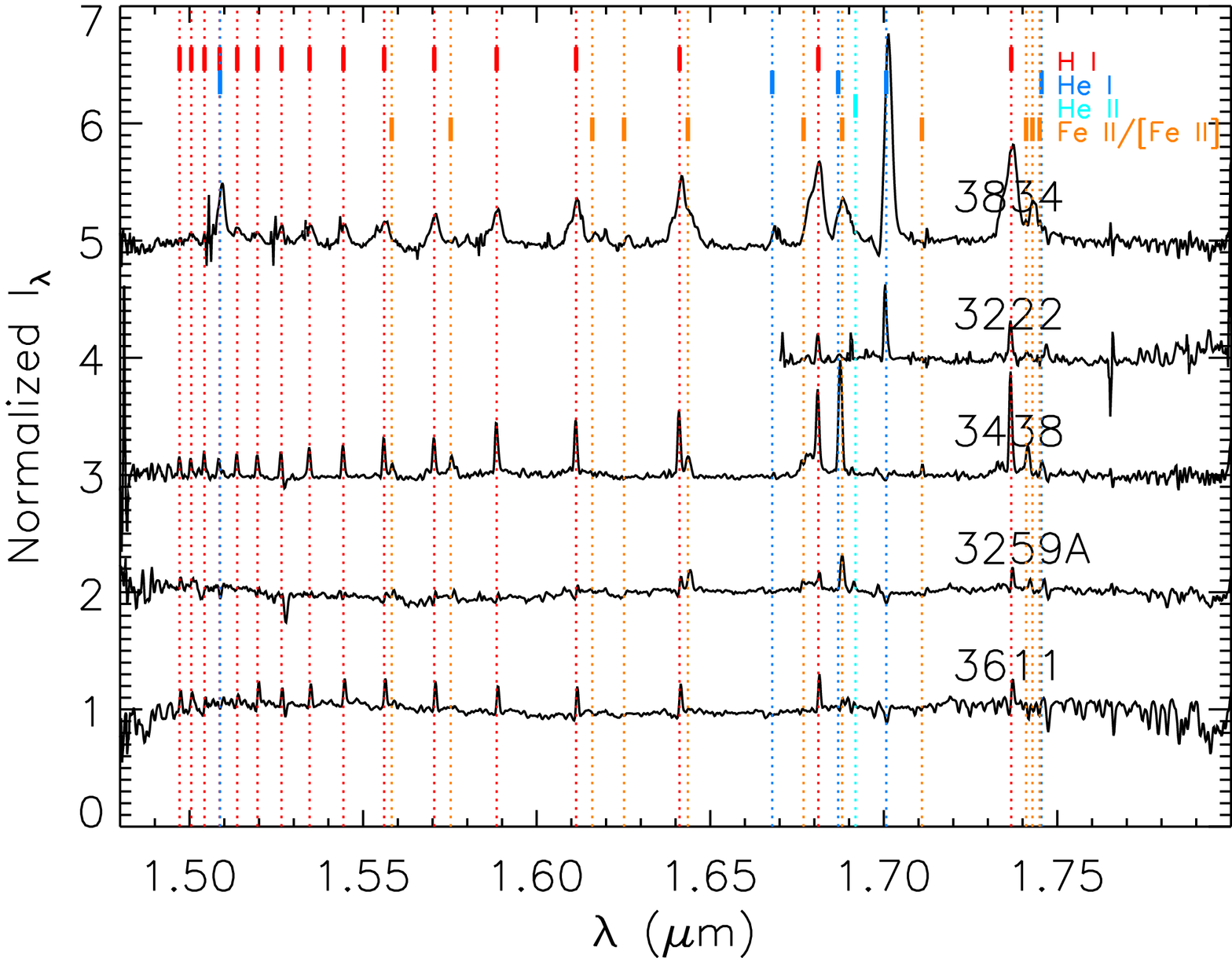}}
  \subfigure[]
  {\label{fig:earlytype_K}
    \includegraphics[angle=90,width=.45\linewidth]{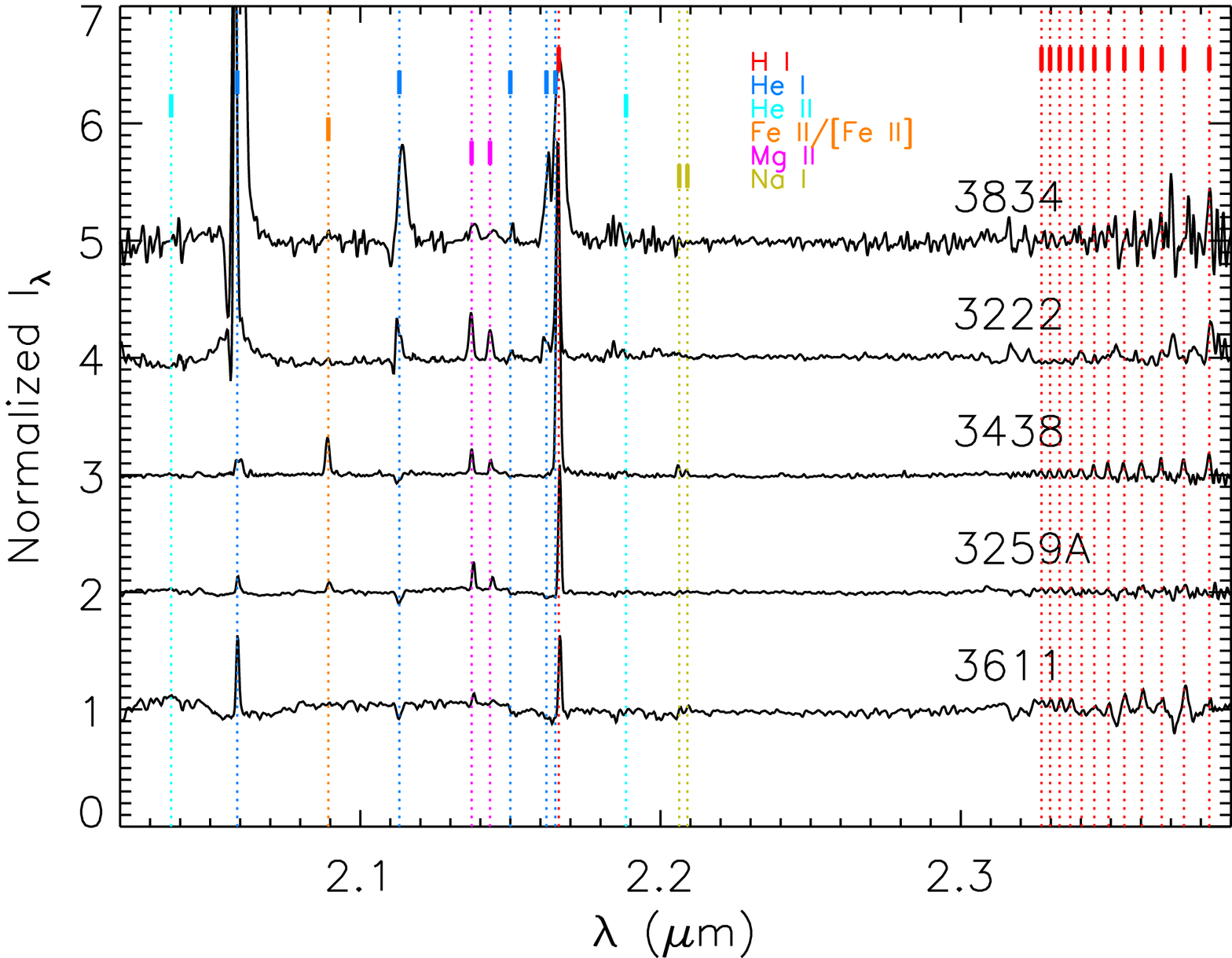}}
  \caption{Near-IR spectra (a: {\it J} band, b: {\it H} band, c: {\it K} band) of the emission line stars that we suggest as LBV candidates. For the stars at the center of MB3222 and MB3259, the short wavelengths end of the spectra are mostly useless and therefore not shown. The most prominent lines are indicated by vertical dashed lines of different color for different species.}
  \label{fig:lbvtype_spec}
\end{figure*}

\subsubsection{LBV candidates}
\label{sec:lbv}

The spectra of the five LBV candidates are shown in Figure \ref{fig:lbvtype_spec}. In addition to the metal lines listed in Table \ref{tab:early_det}, the spectra also show H and \ion{He}{1} emission lines. Variations, both subtle and significant, especially in terms of the intensities and widths of the H and \ion{He}{1} lines, are found in this group, as \citet{Wachter2010} observed in their group of Be/B[e]/LBV candidates.

{\it MB3438} - The spectrum of the central source shows strong hydrogen lines, including the Pfund series at wavelengths longer than 2.3\mic\ and the Brackett series down to about 1.5\mic, and many metal lines. Helium lines are weak or in absorption (e.g. 1.70, and 2.11\mic). The overall spectrum is nearly identical to those of Be/B[e]/LBV in the Group 2A of \citet{Wachter2010}. \citet{Wachter2011} classified it as a B[e]/LBV star, although no reason was given to rule out the Be interpretation.

{\it MB3259} - The spectrum of the central source in MB3259 is similar to that of the star at the center of MB3438, except that most H lines are significantly weaker, especially in the {\it H} band, although iron lines are relatively stronger.

{\it MB3611} - The spectrum of the central source in MB3611 is similar to that of the star at the center of MB3438, although no iron lines are detected except perhaps in the {\it J} band at 1.05, 1.09, and 1.25\mic. On the contrary, many H lines are detected in the Brackett series and possibly in the Pfund series. Multiple \ion{He}{1} transitions are seen in absorption in the {\it J} band. \citet{Wachter2011} classified it as a Be star. However, their motivation for such classification was not given.

{\it MB3222} - The star at the center of MB3222 is very faint in the {\it J} band (17.7 mag) and the spectrum we have obtained has an insufficient signal-to-noise ratio to be shown here. The {\it H} band spectrum also suffers from significant residual emission due to the telluric lines and only part of it is shown in Figure \ref{fig:lbvtype_spec}. Compared to the spectrum of the central source in MB3438, that of the star at the center of MB3222 shows no iron lines. The \ion{He}{1} lines in emission (e.g. at 2.112, 1.7\mic) exhibit P Cygni profiles. We follow the same method as for WR149 and infer a terminal velocity of the wind in MB3222 of about 200~km/s, which is in agreement with an LBV nature.

{\it MB3834} - The spectrum of the star at the center of MB3834 shows broad and strong H lines in the Brackett series down to below 1.55\mic\ and strong \ion{He}{1} lines in emission with P Cygni profiles. Iron lines may be detected at 1.64, 1.68, or 1.69\mic, but each may be blended with H or \ion{He}{1} lines. The most plausible detections of iron lines are at 1.74\mic\ and around 1.62\mic. Using the same method as for WR149, we infer a terminal velocity of the wind in MB3834 of about 550~km/s, which is in agreement with an LBV nature. \citet{Wachter2011} classified it as a B[e]/LBV star, although no reason was given to rule out the Be interpretation.\\

In summary of section \ref{sec:emi}, we confirm in the near-IR the spectral types of WR149 and WR138a and three LBV candidates, previously obtained in the optical domain and, among the MBs, identify a WC5-6 star possibly of low mass, a candidate O5-6 star possibly dwarf, and two Be/B[e]/LBV candidates. As pointed out by \citet{Morris1996} and \citet{Wachter2010}, the {\it JHK} spectra of Be, B[e], and LBV stars are almost identical, although the natures of the objects are quite different. Detailed discussions about Be stars can be found in \citet{Jaschek1981}, \citet{Collins1987}, and \citet{Porter2003}, about the B[e] phenomenon in \citet{Lamers1998} and \citet{Bjorkman1998}, and about LBVs in \citet{Humphreys1994}, \citet{Nota1995} and \citet{Clark2005}. \citet{Lenorzer2002} has established a diagnostic, based on hydrogen lines in the {\it L} band, to distinguish between Be stars, B[e] phenomena, and LBVs. Spectroscopic observations in the 3-4~$\mu$m range of the many LBV candidates among the central sources of MBs could thus remove the ambiguity inherent to their {\it JHK} spectra.

\subsection{Detections of near-IR shells}
\label{sec:shell}

While the main goal of our observations is to identify the spectral types of the stars at the centers of the MBs, we also report for the first time the discovery of several shells in the near-IR spectroscopic observations. They appear very clearly in the differences between the two nodding positions. The MBs with detected shells are MB3214, MB3367, and MB3438. However, in MB3438, the compact extended emission is difficult to assess due to the intensity of the central point source. The lines for which we claim a clear detection of extended emission are the Pa$\delta$~1.006, \ion{He}{2}~1.012, \ion{He}{1}~1.083, Pa$\gamma$~1.095, \ion{He}{2}~1.163, Pa$\beta$~1.283, Br(11-4)~1.682, Br(10-4)~1.737, \ion{He}{1}~2.060, Br$\gamma$~2.168, and \ion{He}{2}~2.189\mic. These lines are all detected in MB3214 while only some are in MB3367.

\begin{figure*}[!t]
  \centering
  \subfigure[]
  {\label{fig:}
    \includegraphics[angle=90,width=.475\linewidth]{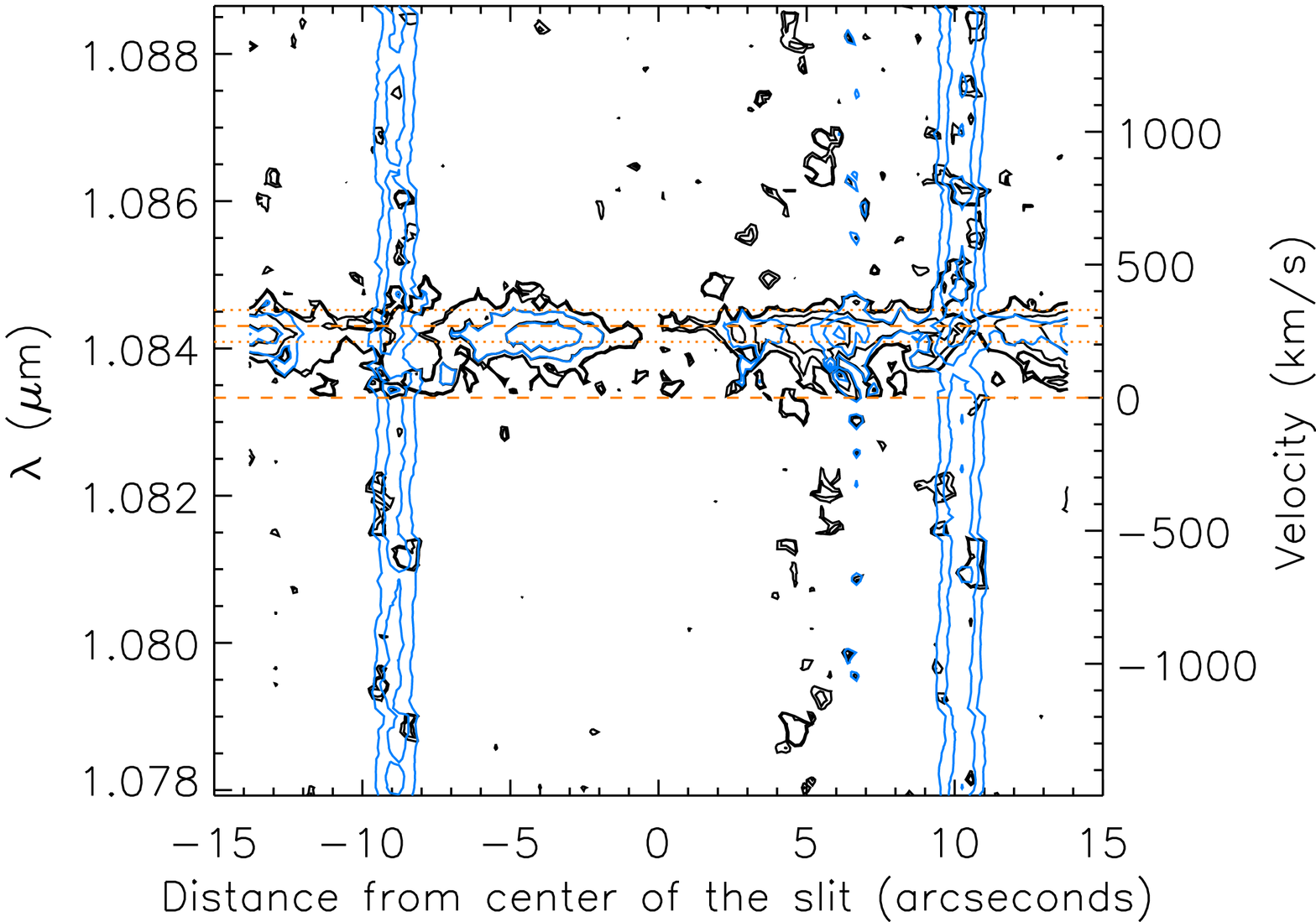}}
  \subfigure[]
  {\label{fig:}
    \includegraphics[angle=90,width=.475\linewidth]{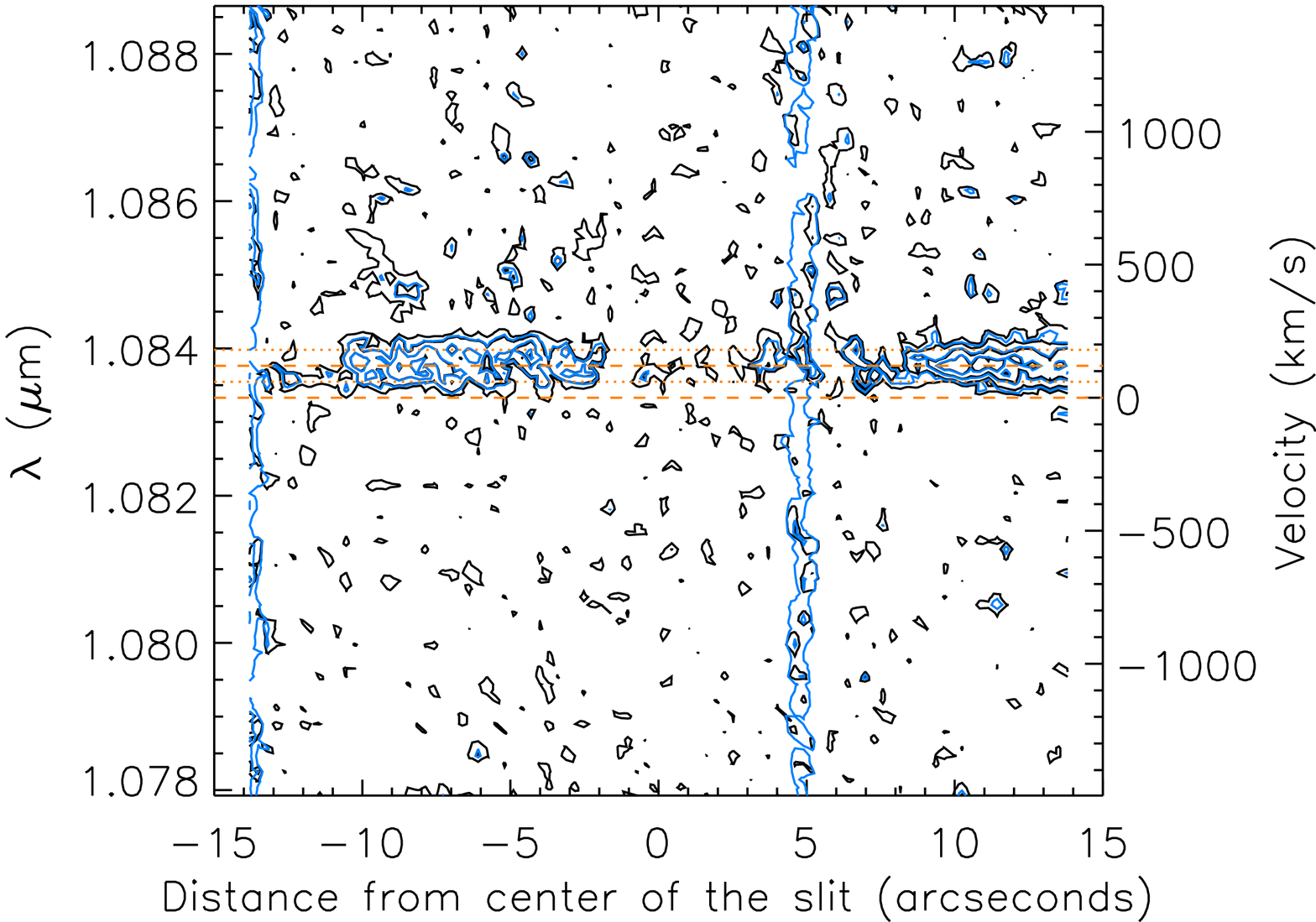}}
  \subfigure[]
  {\label{fig:}
    \includegraphics[angle=90,width=.475\linewidth]{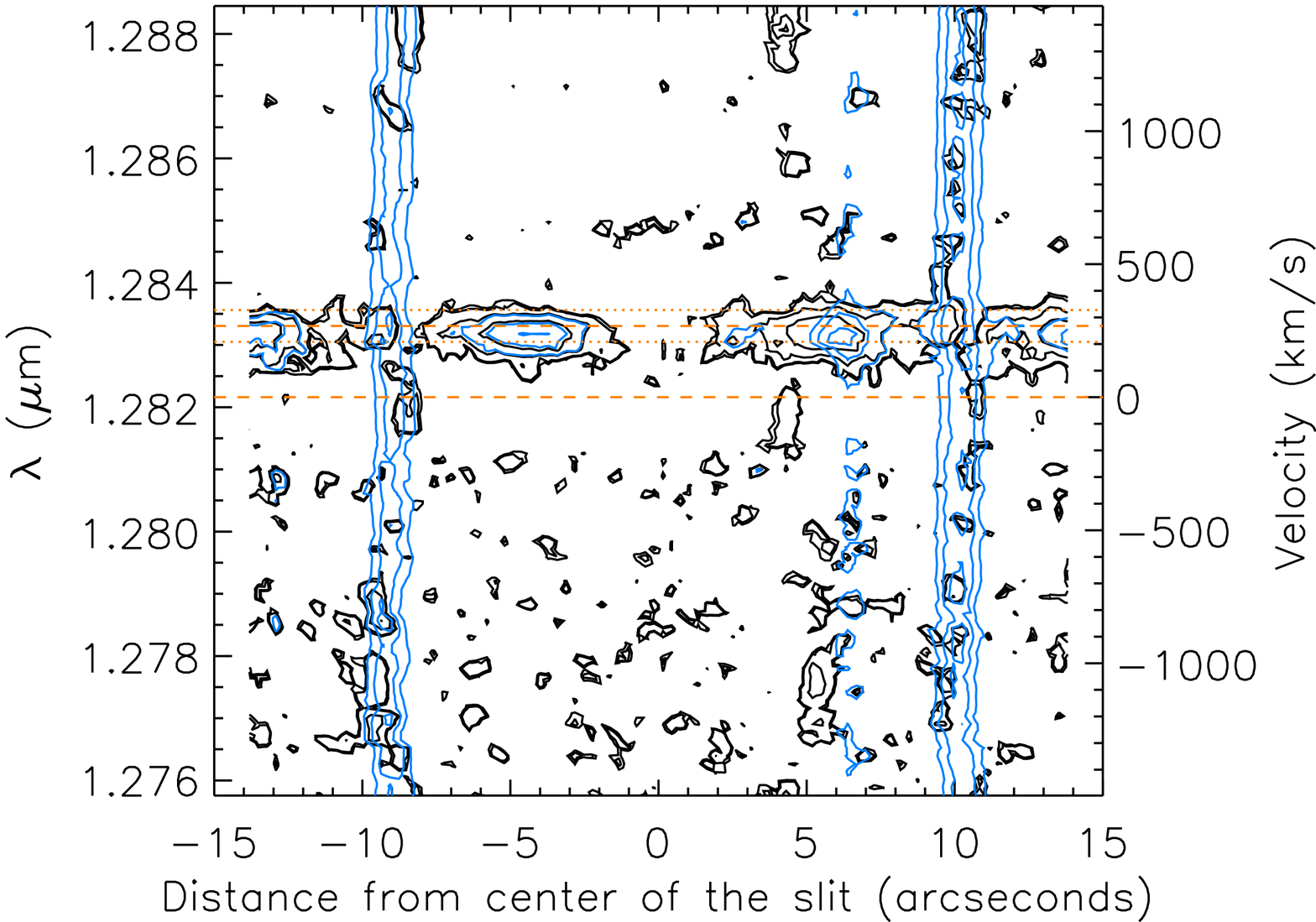}}
  \subfigure[]
  {\label{fig:}
    \includegraphics[angle=90,width=.475\linewidth]{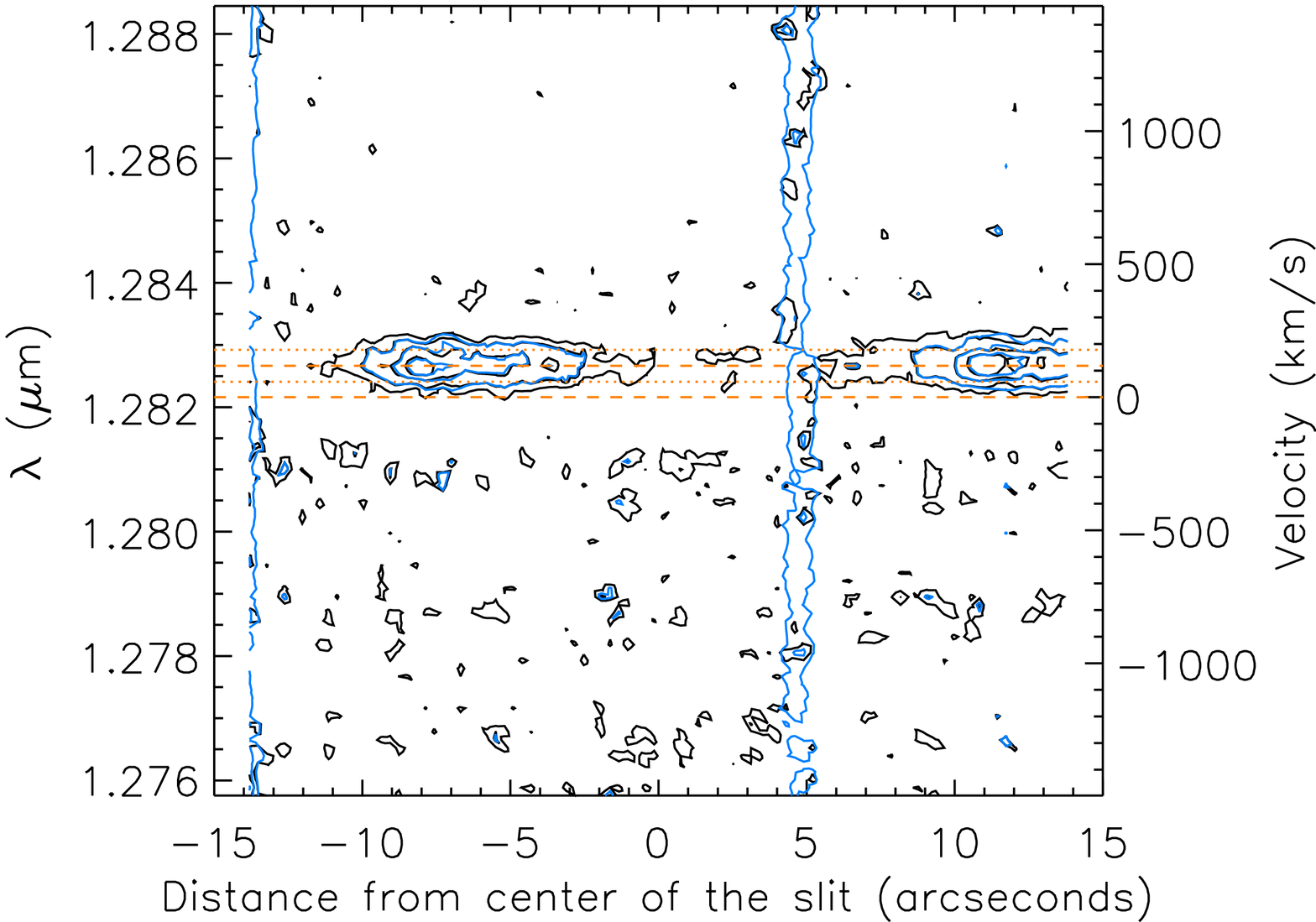}}
  \subfigure[]
  {\label{fig:}
    \includegraphics[angle=90,width=.475\linewidth]{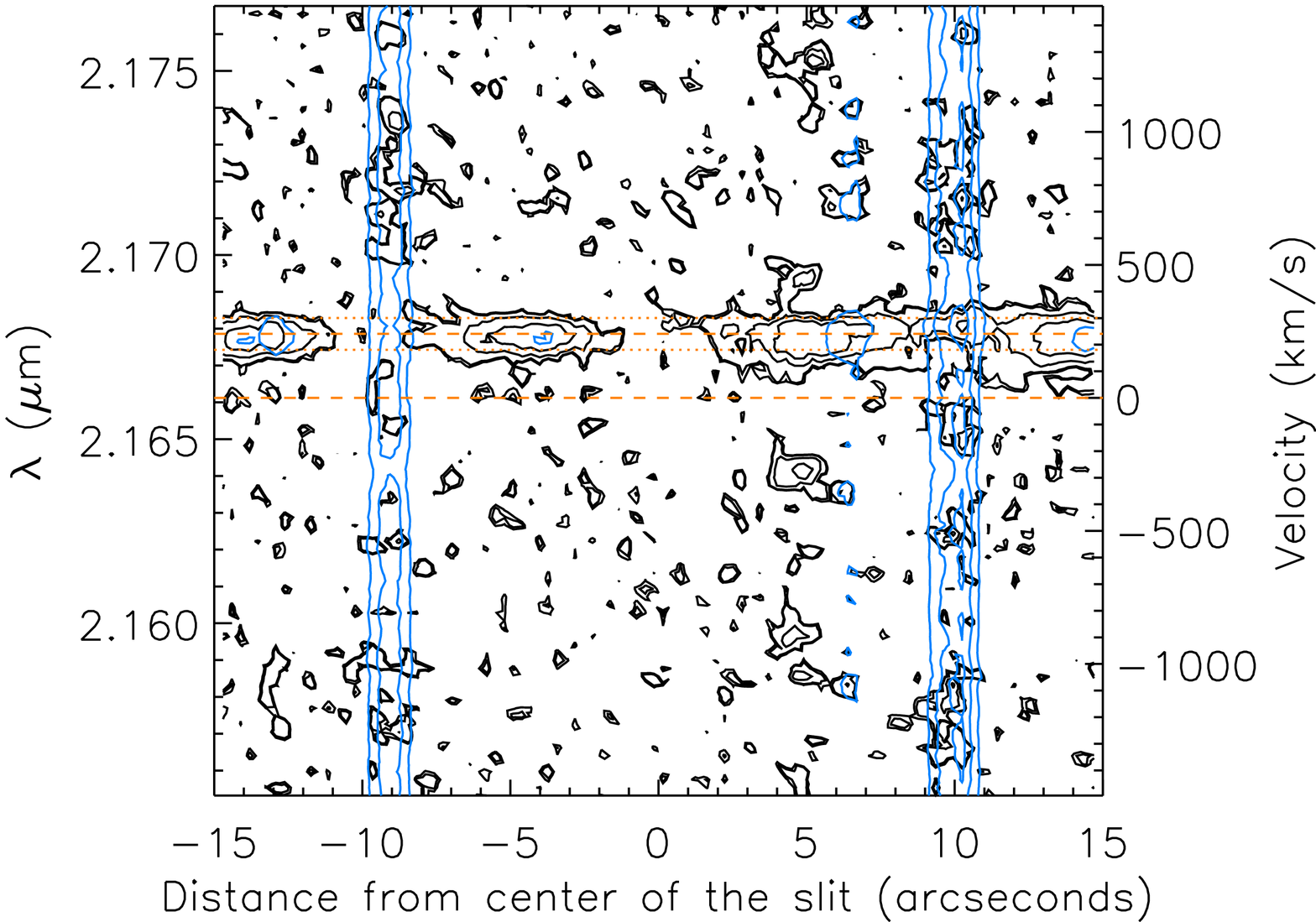}}
  \subfigure[]
  {\label{fig:}
    \includegraphics[angle=90,width=.475\linewidth]{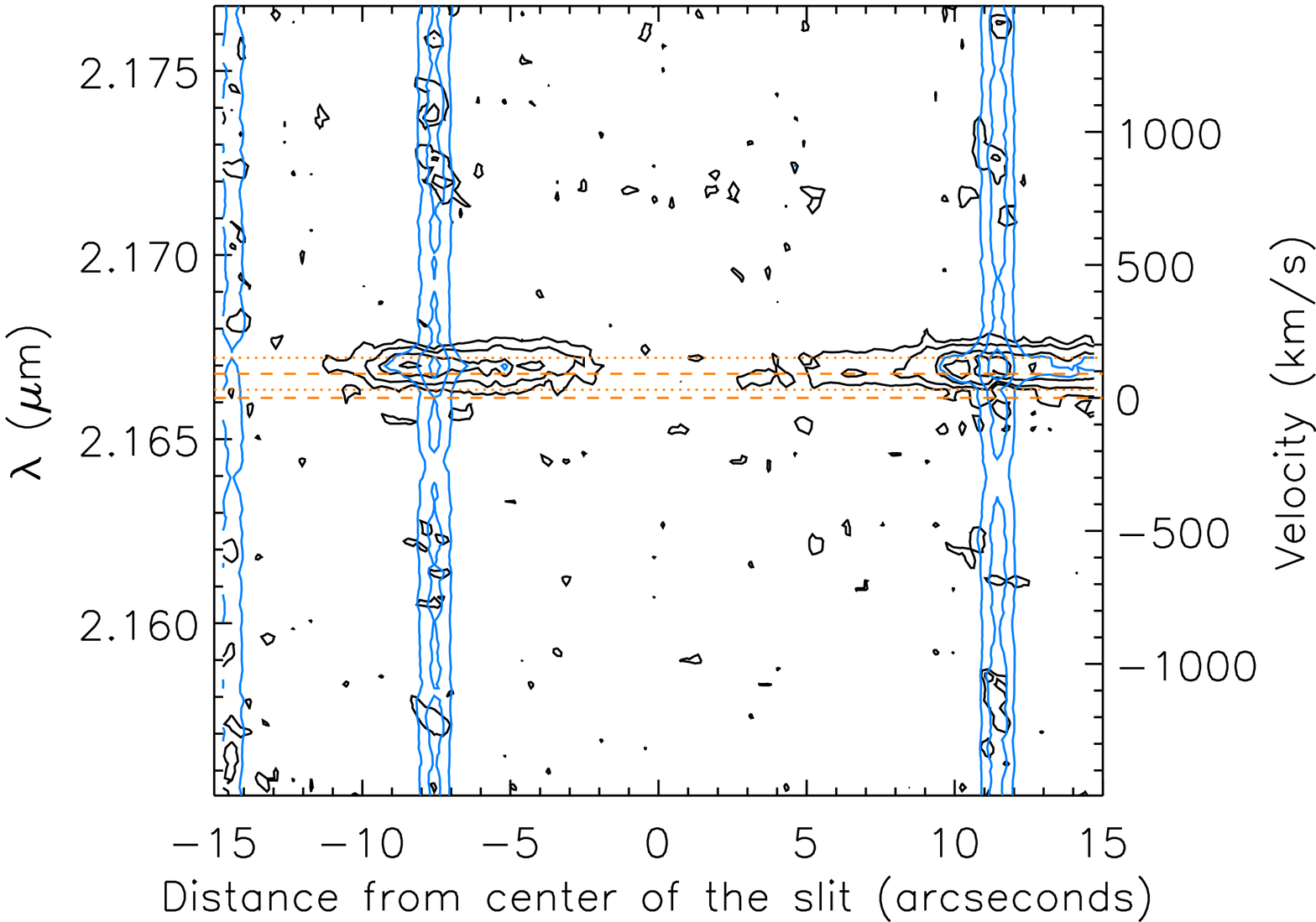}}
  \caption{Position-lambda or position-velocity diagrams for the \ion{He}{1}~1.083\mic, Pa$\beta$, and Br$\gamma$ lines ({\it from top to bottom}) towards MB3214 ({\it left}) and MB3367 ({\it right}). In each panel, NodA is on the right of the slit while NodB is on the left. Contours are at 7.5, 10, 25, and 50\% of the peak intensity in the spectral and spatial range shown, after background (star and atmosphere) subtraction for MB3214, and 10, 25, 50 ad 75\% for MB3367 . Contours in blue are inferred the same way but before subtraction of the star contribution to highlight the position of the central source, at about -10\arcsec\ and +10\arcsec. In the case of MB3367, the source is very faint in J and H bands and the blue contours show the position of another star in the slit. Dashed orange lines show the tabulated and observed wavelengths of the lines. Dotted orange lines show the extend of the resolution width. See text for details.}
  \label{fig:shell}
\end{figure*}

To measure the spatial and velocity extent of a detected shell, we first ``rectified'' the 2D image from TripleSpec using an IDL code written by P.~Muirhead and J.~Wright\footnote{\url{http://astrosun2.astro.cornell.edu/~muirhead/tspec_rectify.pro}}. All the orders of the spectrograph are concatenated and their curvatures corrected so the data appear as a long 2D array, where one dimension indicates the wavelength while the other indicates the position along the slit. We do this on the difference between the two nodding positions, which allows us to subtract the continua from the stars in the slit (by removing the median value along each spatial column of pixels), and the atmospheric lines (by removing the median value along each spectral row of pixels), and thereby to produce position-wavelength and position-velocity diagrams of the shells. Figure \ref{fig:shell} shows such diagrams for the \ion{He}{1}~1.083\mic, Pa$\beta$, and Br$\gamma$ lines in MB3214 and MB3367. Both nodding positions are visible, each on one side of the slit. We note a slight but systematic shift in the wavelength position of the lines towards both targets. For MB3214, the shift is about 250~km/s while it is about 100~km/s for MB3367. Part of this shift could be due to wavelength calibration errors. However, the fact that the shift is systematic indicates that it is real. Using a Galactic rotation curve with $V_{sun} = 254$~km/s and $R_{sun} = 8.4$~kpc \citep{Reid2009}, we find that toward MB3214 ($l\sim20^\circ$), $v_{LSR}$ could reach values higher than 150~km/s between 7 and 9~kpc from the Sun, which is in agreement with our estimates based on the spectral type of the star in MB3214 (11.8$\pm$4.1~kpc, see section \ref{sec:late}). Toward MB3367 ($l\sim32^\circ$), $v_{LSR}$ reaches 100~km/s or above between 5 and 10~kpc from the Sun, in fair agreement with the 11~kpc we suggest if MB3367 has an O5-6~V star at its center (see section \ref{sec:wr}). However, because the MBs appear to be isolated, chances are they are runaway stars \citep[e.g.][]{Gvaramadze2012b}: their radial velocities could thus be significantly different from the average motion of the Galaxy in their vicinity.

From the position-velocity diagrams, we also infer that the shells are barely resolved. Toward MB3214, we measure a $\lambda/FWHM$ of 2400, 2200, and 2400 at $\lambda=$1.083, 1.283, and 2.168\mic\ respectively, which correspond to about 130~km/s. Toward MB3367, those values are 2200, 2400, and 2600 respectively. These values are very close to the resolution of TripleSpec. We therefore cannot rule out that there is some intrinsic broadening of the lines. Along the spatial dimension, we first note that the shell of MB3214 shows a ``lobe'' on each side of the central star, extending out to about 9\arcsec\ with a ``hole'' of about 1-2\arcsec\ radius near the source. The shell of MB3367 shows a single structure that extends about 5\arcsec\ from the central source. In the MIPS24~$\mu$m images, neither object shows a central ``hole'', and we measure a radius of 10-15\arcsec\ for MB3214 and of 7-11\arcsec\ for MB3367. However, the angular resolution of the MIPS24~$\mu$m images is 6\arcsec. Therefore, the radius of the shell measured in the near-IR seems in good agreement with that derived from the mid-IR images. The emission lines observed in the near-IR thus arise from the same region that dominates the mid-IR emission. \citet{Flagey2011b} and \citet{Nowak2014} have shown that they can be accounted for by highly ionized gas emission, warm dust, or a mix of both.

\section{Summary of the findings on the MBs}
\label{sec:sum}

When \citet{Mizuno2010} published their catalog, about 85\% of the MBs were completely ``unknown'' (i.e. they had never been detected in previous observations or no spectral type had been suggested for their central star, if any has been detected). Among the ``known'' MBs, about 80\% were identified as PNe while the others comprised supernova remnants, luminous blue variables and other emission line stars. The PNe especially dominated the group of MBs for which no central star was detected at 24~$\mu$m. About four years later, combining the results from \citet{Wachter2010}, \citet{Gvaramadze2010}, \citet{Mauerhan2011}, \citet{Wachter2011}, \citet{Flagey2011b}, \citet{Nowak2014}, and the present paper, the number of identified MBs has more than doubled, with a total of 128 objects have at least suggested classifications. The fraction of massive evolved stars (candidates) is about half of the ``identified'' MBs. Table \ref{tab:sum} summarizes the statistics of those findings.

\begin{table*}[t]
  \centering
  \begin{tabular}{c c c c c c c c c c}
    \hline
    Nature & Number & Comments \\
    \hline
    \hline
    GK (super)giants & 7 & \\
    \hline
    F/G & 1 & \\
    \hline
    B/A & 2 & \\
    \hline
    B stars & 5 & from B0 to B9 \\
    \hline
    Be, B[e], and/or LBV & 30 & Only few are known, {\it bona fide} LBV \\
    \hline
    OB & 5 \\
    \hline
    Oe/WN & 2 & \\
    \hline
    WR & 13 & 11 WN, 1 WC or [WC], 1 [WO] \\
    \hline
    Galaxies & 2 & \\
    \hline
    PNe & 58 & Most central sources not characterized \\
    \hline
    SNR & 3 & \\
    \hline
    Unknown & 300 \\
    \hline
  \end{tabular}
  \caption{Summary of the identifications of MBs in \citet{Mizuno2010}.}
  \label{tab:sum}
\end{table*}

Among the 62 shells selected by \citet{Wachter2010}, 48 are part of the \citet{Mizuno2010} catalog and all except seven central sources have now been identified. \citet{Gvaramadze2010} established a list of 115 nebulae from their analysis of the first part of the MIPSGAL survey (i.e. $10^\circ<|l|<65^\circ$) and the Cygnus-X survey. There are 60 MBs in their catalog, 29 of which were not already in the list of \citet{Wachter2010}. Among these 29 MBs, 16 remain to be identified. \citet{Mauerhan2011} found a WN8-9h star at the center of one MB, which was also part of the sample of four MBs observed with the high-resolution module of {\it Spitzer}/IRS by \citet{Flagey2011b}. \citet{Stringfellow2012b} independently confirmed this star as a WN7-9h. While most of the new identifications have been obtained from optical and near-IR spectra of the central sources, some have also been suggested based on mid-IR spectroscopic observations with {\it Spitzer}/IRS of the shells, and their central sources when present \citep{Flagey2011b, Nowak2014}. Those unique datasets have revealed a population of highly excited, dust-poor PNe, at the centers of which some types of extreme white dwarf may be hidden, accounting for 9 out of the 14 MBs in this sample. The mid-IR observations have also made possible the characterization of the dust properties in the nebula of 5 dust-rich MBs. These last two studies, although based upon a small number of MBs, indicate that the dust-rich spectra are associated with massive star candidates (e.g. LBV, WR) that are detected in the mid-IR, while the highly excited, dust-poor MBs show no central source in this wavelength range.

Among the 128 ``known'' MBs, the PN candidates now constitute less than 45\%. Most of these have central sources that have not been characterized. Their true natures could thus be different. For instance, MGE333.9202-00.8910 was originally associated with a PN in \citet{Mizuno2010} but \citet{Wachter2011} identified the central source as a B[e]/LBV candidate. The group of Be/B[e]/LBV accounts for almost half of the remaining ``known'' objects (30/70). Only a few {\it bona fide} LBV, or LBV candidates suggested by other means, are among this group, including V~481~Sct (MB3280), and Wray~17-96 (MB4562). The ambiguity between the Be, B[e], and LBV phenomena is inherent to near-IR spectroscopic observations of these sources, and only further observations will help us distinguish between the three types (see section \ref{sec:lbv}). As only a few {\it bona fide} LBV are known \citep{Clark2005}, and as they are among the most massive, bright, and extreme stars in terms of mass loss events, the finding of several tens of new candidates could be a major step towards a better understanding of this critical step in the evolution of massive stars toward the WR and SN phases \citep[see e.g.][]{Groh2013}.

The identifications of the central sources in the MBs have also revealed 10 newly discovered WR stars, in addition to the WN5b found via {\it J}, {\it K} and narrowband imaging surveys of the Galactic plane \citep[MB4107,][]{Shara2009}. A low-mass [WO~3]pec star \citep[at the center of MB4344,][]{Weidmann2008} was also known before the publication of the \citet{Mizuno2010} catalog, and a possible [WC5-6] is identified in the present paper. A close pair of WN8-9h (MB3312, MB3313), located at about 5~kpc from the Sun, is among the most striking discoveries, as both mid-IR shells seem to be interacting \citep{Mauerhan2010, Gvaramadze2010, Wachter2010, Burgemeister2013} and offers an unprecedented view of interacting winds, outflows, and radiations of evolved massive stars. The search for Galactic WR stars has been very active in the recent years as models predict there are ten times as many as are known. The finding here of 10 new Population~I WR stars is not a significant contribution to this search, and the detection rate is not competitive with other methods \citep{Mauerhan2011, Shara2009, Shara2012, Faherty2014,Kanarek2014x}. However, the use of the mid-IR as a selection criteria, via the detection of the nebula, could facilitate the discovery of the WR stars hidden by the interstellar extinction in the Galactic plane at shorter wavelengths.

Massive star candidates dominate the new discoveries, i.e. after the publication of the \citet{Gvaramadze2010}, \citet{Wachter2010}, and \citet{Mizuno2010} catalogs. This is because evolved massive stars are more likely to be surrounded by hot or warm dust, which increases their near- to mid-IR luminosities, while hot, low luminosity white dwarfs remain hidden, especially behind the significant interstellar extinction of the Galactic plane. The sample presented in this paper corresponds to average magnitudes of 11.5, 12.6, and 14.2 in {\it K}, {\it H} and {\it J} bands, respectively. At these magnitudes, and even more at deeper levels, confusion between multiple potential central sources is an issue. Although the UKIDSS survey goes a few magnitudes deeper than 2MASS \citep[{\it K}$\sim$18~mag, ][]{Lawrence2007}, about one-third of the MBs still have no clear candidate central sources. On the other hand, about one-third of the MBs has at least two sources within 2\arcsec\ of their geometric center. It will be necessary to use different observational methods to determine the true natures of the remaining $\sim$300 unknown MBs. Because of the significant extinction within the Galactic plane, these observations will surely be limited to the IR wavelength range. Near-IR integral field instrument, such as SINFONI on the VLT or NIFS on Gemini-North, would allow one obtain the spectra of multiple candidate central sources simultaneously.

\section{Conclusion}
\label{sec:ccl}

We have presented near-IR spectroscopic observations of the central source in 14 circumstellar shells, previously discovered in the {\it Spitzer}/MIPSGAL 24~$\mu$m images of the Galactic plane survey, as well as 3 comparison sources. The near-IR TripleSpec data allowed us to identify:

\begin{itemize}
\item 5 late type giants, thanks to the detection of the CO band-head absorption features. We used the equivalent width of the 2.29~$\mu$m feature to derive two possibles spectral type (one for each of the giant and supergiant class) and ruled out the supergiant interpretation using the inferred distance and total extinction along the line of sight.

\item 3 early type stars (B and A), likely supergiants or giants, identified by the strengths and widths of the H and He absorption features. In particular, we revoked the identification of BD+43~3710 as a carbon star, and suggested its spectral class is B5~I, while the true carbon star is the nearby ($\sim4$\arcmin) V2040 Cyg. We used their magnitudes and colors to set constraints on their distances and the sizes of the mid-IR nebulae.

\item 4 O and WR stars, characterized by the \ion{He}{2} line at 2.1885\mic. We used diagnostics based on line equivalent width ratios and spectral libraries to infer the spectral types of two MBs to be WC5-6, possibly of low mass, and a candidate O5-6, possibly dwarf. The [WC5-6] is a confirmation and refinement of the suggested identification by \citep{Gvaramadze2010}. The two other WR stars are among the three comparison sources. Their spectral types (WN9h and a WN5sh) are in excellent agreement with those derived from optical observations. The three WR stars also exhibit typical broad lines with terminal velocities of 650 and 1700~km/s for the WN9h and WN5sh respectively, and a FWHM of about 1250~km/s for the WC5-6.

\item 5 LBV candidates whose near-IR spectra are similar to that of Be or B[e] stars, as shown in several other studies. We followed the notation of \citet{Wachter2010} and labelled them Be/B[e]/LBV. For two of these objects we estimated terminal velocities of 200 and 550~km/s for their stellar winds. For three of these LBV candidates, we confirm their status previously suggested by \citet{Wachter2011} and \citet{Stringfellow2012b}.

\end{itemize}

The long-slit Palomar/TripleSpec observations have also revealed at least two shells in emission in several H and He lines. The regions traced by the extended line emission have spatial extents in very good agreement with those seen in the mid-IR images. The spreads in velocity of the shells are $\lesssim150$~km/s. Constraints on the distances to the MBs will allow us to derive, in a future paper, the dust masses in the nebulae, for those that are detected in the Herschel far-IR bands \citep{Flagey2014b}.

Near-IR and optical long-slit spectroscopic observations have now doubled the number of MBs for which natures have been identified or suggested. They have revealed many WR stars and LBV candidates, reducing the fraction of PNe from about 80\% when the catalog of \citet{Mizuno2010} was published, to less than 50\%. However, the new identifications suffer from selection effects, as these central sources are the brightest of the 428 MBs in the near- to mid-IR, with typical {\it K} magnitudes of about 10.5. Because most of the remaining 300 unidentified MBs are within 1$^\circ$ of the Galactic plane and toward the inner Galaxy (63\% are within $|l|<10^\circ$), confusion and extinction are limiting factors that will require the use of near-IR integral field units (e.g. SINFONI on VLT, NIFS on Gemini-North) rather than slit spectrographs to identify the natures of the potential central sources. Deep imaging will also be necessary to find the central sources in the MBs where none are detected so far. The characterization of the mid-IR extended emission of the MBs, which started on a limited sample at the end of {\it Spitzer}'s cryo-mission, has set constraints on the dust masses and mass loss rates in dust-rich MBs, and revealed peculiar, highly excited, dust-poor objects that may harbor extreme white dwarfs \citep{Flagey2011b, Nowak2014}. Additional such observations will have to wait for the next space mission with mid-IR spectroscopic capabilities: the James Webb Space Telescope.

\acknowledgements The Hale Telescope at Palomar Observatory is operated as part of a collaborative agreement between the California Institute of Technology, its divisions Caltech Optical Observatories and the Jet Propulsion Laboratory (operated for NASA), and Cornell University. This research has made use of the SIMBAD database, operated at CDS, Strasbourg, France, and of the NASA/ IPAC Infrared Science Archive, which is operated by the Jet Propulsion Laboratory, California Institute of Technology, under contract with the National Aeronautics and Space Administration. NF thanks Francois Ochsenbein for providing the plate scale of the BD charts, Paul Crowther for his help with some identifications, and the referee for very valuable comments. TRG’s research is supported by the Gemini Observatory, which is operated by the Association of Universities for Research in Astronomy, Inc., on behalf of the international Gemini partnership of Argentina, Australia, Brazil, Canada, Chile, and the United States of America.

\bibliographystyle{aa} 
\bibliography{../../nicolasflagey} 

\end{document}